\documentclass{emulateapj}
\usepackage{graphicx}
\usepackage{amssymb}
\usepackage{amsmath}
\usepackage{subfigure}
\usepackage{multirow}
\usepackage{threeparttable}
\usepackage{array}
\usepackage{natbib}
\usepackage{xcolor}
\usepackage{subfigure}
\usepackage{hyperref}

\hypersetup{
    colorlinks=true,
    linkcolor=red,
    filecolor=magenta,
    urlcolor=cyan,
}

\def\kms{~\rm km~s^{-1}}
\def\cmsq{~\rm cm^{-2}}
\def\cc{~\rm cm^{-3}}

\begin{document}
\title{{\it HST}/COS Observations of the Warm Ionized Gaseous Halo of NGC 891}
\author{Zhijie Qu, Joel N. Bregman, Edmund J. Hodges-Kluck}
\affil{Department of Astronomy, University of Michigan, Ann Arbor, MI 48104, USA}
\email{quzhijie@umich.edu}

\begin{abstract}
The metallicity of galactic gaseous halos provides insights into accretion and feedback of galaxies.
The nearby edge-on galaxy NGC 891 has a multi-component gaseous halo and a background AGN (LQAC 035+042 003) projected 5 kpc above the disk near the minor axis.
Against the UV continuum of this AGN, we detect lines from 13 ions associated with NGC 891 in new {\it HST}/COS spectra.
Most of the absorption is from the warm ionized gas with $\log T=4.22\pm0.04$, $\log n_{\rm H}=-1.26\pm0.51$, and $\log N_{\rm H}=20.81\pm 0.20$.
The metallicity of volatile elements (i.e., C, N, and S) is about half solar ($\rm[X/H] \approx -0.3\pm 0.3$), while Mg, Fe, and Ni show lower metallicities of $\rm[X/H]\approx-0.9$. 
The absorption system shows the depletion pattern seen for warm Galactic diffuse clouds, which is consistent with a mixture of ejected solar metallicity disk gases and the hot X-ray emitting halo ($Z=0.1-0.2Z_\odot$).
The warm ionized gases are about 5 times more massive than the cold \ion{H}{1} emitting gases around the galactic center, which might lead to accretion with a mean rate of $10^2~M_\odot\rm~yr^{-1}$ for a period of time.
We also detect low metallicity ($\approx 0.1~Z_\odot$) gases toward LQAC 035+042 003 at $110\kms$ (a high velocity cloud) and toward another sight line (3C 66A; 108 kpc projected from NGC 891) at $30\kms$.
This low metallicity material could be the cold mode accretion from IGM or the tidal disruption of satellites in the NGC 891 halo.
\end{abstract}

\maketitle

\section{Introduction}
The formation of $L^*$ galaxies leads to a hot halo with $T \sim T_{\rm virial}$ and extending to or beyond the virial radius, ($R_{\rm vir}$; \citealt{White:1978aa, Cen:1999aa, Mo:2010aa, Sokoowska:2018aa}).  The gas density of the hot halo increases inward so that the cooling time falls below the Hubble time at $R \sim 100$ kpc \citep{Thompson:2016aa, Qu:2018aa}.  As this gas cools, it slowly flows inward and can accrete onto the galaxy, provided that feedback from the galaxy does not completely offset the cooling \citep{Keres:2005aa, Nelson:2013aa}.

Feedback, at the level we expect today in spiral galaxies, will expel hot gas from the disk but rarely at velocities great enough to unbind the gas from the galaxy halo, even during starburst events \citep{Veilleux:2005aa, Fielding:2017aa}.  This feedback creates a galactic fountain of typical height $5-10$ kpc above the disk and is seen in X-ray emission studies of edge-on galaxies (e.g., \citealt{Strickland:2004aa, Hodges-Kluck:2013aa, Li:2013aa}).  One does not expect a sharp boundary between the galactic fountain and extended hot halo, as particularly hot plumes can rise further and there can be mixing between the two regions (e.g., \citealt{Fielding:2017aa}).  Nevertheless, regions close to the disk should be dominated by galactic fountain activities and regions further away by the long-lived hot halo.

Metallicity may offer a diagnostic to distinguish between the extraplanar and halo gas.
The metallicity of galactic fountain gas should be about solar \citep{Fox:2015aa, Fox:2016aa, Savage:2017aa}, as it flows upward from the disk, although the metallicity may be further enhanced by the supernovae that drive the flow \citep{Scannapieco:2008aa, Li:2017ab}.
The metallicity of the extended hot halo gas should be lower, as material falling onto galaxies is predicted to have less enrichment \citep{Oppenheimer:2016aa}.
This is supported by absorption line observations of extended gas, where the metallicity is typically $0.1-0.5 Z_\odot$ \citep{Lehner:2013aa, Wotta:2016aa, Prochaska:2017aa}.

It is challenging to obtain metallicities near the transition between the galactic fountain and the extended hot halo.  For the Milky Way, the intermediate velocity clouds are typically a few kpc above the disk and from ultraviolet (UV) absorption line studies, one finds a typical metallicity of $\approx 0.5-1.0 Z_\odot$ \citep{Fox:2016aa, Savage:2017aa}, although there are infalling high velocity clouds at heights of $\sim 10\rm~ kpc$ and with metallicities of about $0.3 Z_\odot$ \citep{Wakker:2001aa, Wakker:2008aa, 
Barger:2012aa, Fox:2018aa}.  X-ray emission and absorption studies infer a mean metallicity of about $0.5 Z_\odot$ for gas within 50 kpc \citep{Bregman:2018aa}.

Background AGN sightlines through external galaxies rarely pass through the disk, and when this occurs, it is impossible to spatially assign a height to the absorption.  The exception to this is for edge-on galaxies, which have been studied in X-ray emission \citep{Strickland:2004aa, Hodges-Kluck:2013aa, Li:2013aa} and through dust-scattering halos \citep{Hodges-Kluck:2014aa, Hodges-Kluck:2016ab}.  The dust-scattering halos indicate a change in the nature of the dust metallicity at $5-10$ kpc, suggesting a transition to the extended hot halo. The X-ray emission has the potential to yield the metallicity as a function of height, but this demands lengthy observations with current instruments.  The first observation capable of making this determination, from XMM-Newton observations of NGC 891, is discussed in a separate work \citep{Hodges-Kluck:2017aa, Hodges-Kluck:2018aa}. 

An absorption line study of this transition gas in an edge-on galaxy is possible for NGC 891, where a background AGN (LQAC 035+042 003) projects to 5 kpc above the disk and for the inner part of the galaxy \citep{Bregman:2013aa}.  Spectra taken with STIS on {\it Hubble Space Telescope} ({\it HST}) revealed absorption lines of \ion{Fe}{2}, \ion{Mg}{1} and \ion{Mg}{2}, with different abundances from subsolar to nearly solar.  The limitations of this study are that the lines were not well resolved and the elements probed are subject to depletion onto grains.  These shortcomings could be overcome by the Cosmic Origin Spectrograph (COS; \citealt{Green:2012aa}), which provides high resolution ultraviolet (UV) spectra.

In this paper, we present the analyses of the COS/far-ultraviolet (FUV) spectrum of LQAC 035+042 003, which probes absorption by a number of elements and ionization states, including refractory elements that are not heavily depleted in halo clouds \citep{Savage:1996aa}.
Another target (3C 66A) projected along the disk major axis at $108~ \rm kpc$ is also presented, which probes the outer region of the halo.
In Section 2, we describe the observation data (UV and \ion{H}{1} 21 cm lines), the data reduction method and the column density measurements.
The photoionization models are described in Section 3 for all detected systems, leading to the hydrogen density and hence the path length of the absorption structures.
We discuss the origin for the detected UV absorption systems in Section 4 and summarize our results in Section 5.

\section{Observations and Data Analysis}
\subsection{LQAC 035+042 003}
\subsubsection{The QSO}
The background QSO located at RA $\rm = 02 h 22 m 24.4 s$, DEC $ = +42^\circ 21' 38''$ (J2000) was discovered as an X-ray source in the {\it ROSAT} image by \citet{Read:1997aa} and identified as a QSO in a followup optical spectroscopic observation (see details in \citealt{Bregman:2013aa}). This QSO is projected near the edge-on disk of NGC 891 with an angular separation of $106''$, which is $4.7\rm ~kpc$ at the NGC 891 distance of $9.12 \rm~Mpc$ \citep{Tully:2013aa}. The AGN is also projected close to the minor axis of the disk with a separation of $11''$ ($0.5\rm~kpc$; \citealt{Bregman:2013aa}).

\subsubsection{The {\it HST}/COS Spectrum}
The {\it HST}/COS spectra were acquired on 22-24 August 2013 in the FUV band and 22 July 2013 in the near-ultraviolet (NUV) band ({\it HST} proposal GO 12904; PI: Bregman). 
The high resolution FUV spectrum ($R\approx 20000$) was obtained with the grating G130M, and the total exposure time is 23.7 ks. 
The central wavelength in each exposure is $1291~\rm \AA$, and multiple FP-POS positions are employed to reduce the fixed pattern noise. 
The low resolution grating G230L was employed for the NUV spectrum ($R\approx 2500$), and the total exposure time is 10.2 ks.
Similar to the FUV observations, the NUV exposures have different FP-POS positions at the same central wavelength ($2870\rm~\AA$).

\begin{table}
\begin{center}
\caption{The COS/FUV Absorption Line Measurements}
\label{NGC891}
\begin{tabular}{lcrrrrrr}
\hline
\hline
Ion & $\log N$ & $\sigma_{\log N}$ & $b$ & $\sigma_b$ & $v^a$ & $v_{\rm c}^{g}$ & $\sigma_v$\\
& $\cmsq$ & $\rm dex$ & \multicolumn{2}{c}{$\kms$} & \multicolumn{3}{c}{$\kms$}  \\
\hline
\multicolumn{7}{c}{NGC 891 (LQAC 035+042 003)}\\
\hline
   \ion{C}{1}   &  13.76  &  0.17   &  23.5  & 15.7  &  $-34.6$  &  $-16.7$ & 10.0 \\
   \ion{C}{2}   &  15.8  &  $>$   &  $...$  & $...$  &  $-25.0$  &  $-7.1$ & 2.6 \\
   \ion{C}{2}*   &  14.14  &  0.05   &  33.3  &  5.5  &  $-34.2$  &  $-16.3$ & 3.6 \\
   \ion{N}{1}   &  15.71  &  0.10   &  34.1  &  1.7  &  $-24.0$  &  $-6.1$ & 1.3 \\
   \ion{N}{5}   &  13.97  &  0.10   &  42.8  & 14.7  &  $-27.3$  & $ -9.4$ & 9.4 \\
 \ion{Mg}{2}$^f$  &  15.5 &  $<$ &  $...$  & $...$  &   $...$  & $...$ &  $...$ \\
 \ion{Si}{2}   &  14.8  &  $>$   &  $...$  & $...$  &  $-39.1$  &  $-21.2$ & 1.1 \\
\ion{Si}{3}   &  14.7  &  $>$   &  $...$  & $...$  &  $-50.2$  &  $-32.3$ & 5.3 \\
 \ion{Si}{4}$^b$  &  13.89  &  0.08   &  48.1  &  8.6  &  $-69.2$  &  $-51.3$ & 8.3 \\
 \ion{Si}{4}$^b$   &  13.62  &  0.16   &  22.4  &  7.7  &   $-5.4$  &  12.5 & 4.3 \\
 \ion{Si}{4}$^b$   &  12.93  &  0.31   &  32.5  & 29.4  &   75.8  & 21.8 \\
  \ion{P}{2}   &  13.76  &  0.09   &  25.1  &  9.6  &   $-4.1$  & 13.8 & 6.0 \\
  \ion{S}{2}   &  15.33  &  0.02   &  50.9  &  3.2  &  $-25.5$  &  $-7.6$ & 2.2 \\
 \ion{Fe}{2}   &  15.10  &  0.04   &  55.1  &  4.6  &  $-34.1$  & $-16.2$ & 3.2 \\
 \ion{Ni}{2}   &  14.19  &  0.06   &  53.0  & 10.4  &  $-36.1$  &  $-18.2$ & 6.7 \\

\hline
\multicolumn{7}{c}{NGC 891 HVC}\\
\hline

   \ion{C}{1}$^c$   &  13.66  &  0.23   &   7.3  & 14.0  &   81.4  & 99.3 & 5.7 \\
  \ion{C}{2}   &  13.97  &  0.18   &  10.0  &  $...^d$  &  107.1  &  125.0 & 3.1 \\
   \ion{N}{1}$^f$   &  13.8  &  $<$   &  $...$  & $...$  &   $...$  & $...$ &  $...$ \\
   \ion{N}{5}   &  13.47  &  0.15   &  19.7  & 12.8  &   99.6  & 117.5 & 7.1 \\
 \ion{Si}{2}   &  13.50  &  0.08   &  14.8  &  2.1  &   89.4  & 107.3 & 1.5 \\
\ion{Si}{3}$^e$   &  13.1 &  $>$   & $...$   &  $...$  &   89.4  & 107.3 & $...$ \\
\ion{Si}{4}$^f$   &  12.5  &  $<$   &  $...$  & $...$  &   $...$  &  $...$ &  $...$ \\
 \ion{Fe}{2}   &  13.77  &  0.18   &  20.0  &  $...^d$  &  106.8  & 124.7 & 11.0 \\

\hline
\multicolumn{7}{c}{NGC 891 (3C 66A)}\\
\hline
   \ion{C}{2}   &  13.03  &  0.11   &  10.0  &  $...^d$   &  35.0 & 57.8 &  4.0 \\
   \ion{C}{4}   &  13.79  &  0.03   &  23.2  &  2.3  &  31.1  & 53.9 & 1.5 \\
\ion{Si}{2}$^f$   &  11.9  &  $<$   &  $...$  & $...$  &   $...$  &  $...$  &  $...$ \\
\ion{Si}{3}   &  12.70  &  0.05   &  22.8  &  4.3  &  28.2 & 51.0 & 2.6 \\
 \ion{Si}{4}   &  12.45  &  0.10   &  14.8  &  7.3  &  32.4  & 55.2 & 4.2 \\

\hline
\end{tabular}
\end{center}
$^a$ The velocity shift is calculated at $z=0.001761$, which corresponding to the bulk velocity of NGC 891 ($528.3 \kms$).\\
$^b$ \ion{Si}{4} has multiple components for the absorption associated with NGC 891.\\
$^c$ \ion{C}{1} only has one detected line $\lambda 1328.8\rm~\AA$, while another strong line $\lambda 1277.2\rm~\AA$ is in the gap between two segments.\\
$^d$ For these ions, $b$ factors can not be well constrained, so we fixed them in the fitting at values of $10-20\kms$.\\
$^e$ The HVC \ion{Si}{3} is completely blended with the AGN outflow \ion{Ne}{6}, which leads to large uncertainty of the measurements. See text for discussions on this ion.\\
$^f$ These ions have the upper limits calculated from the limiting EW at the $2\sigma$ level.\\
$^g$ The velocity measurements are corrected by aligning the FUV spectra and the \ion{H}{1} 21 cm lines (see the text for details).
\end{table}

\begin{figure*}
\begin{center}
\includegraphics[width=0.96\textwidth]{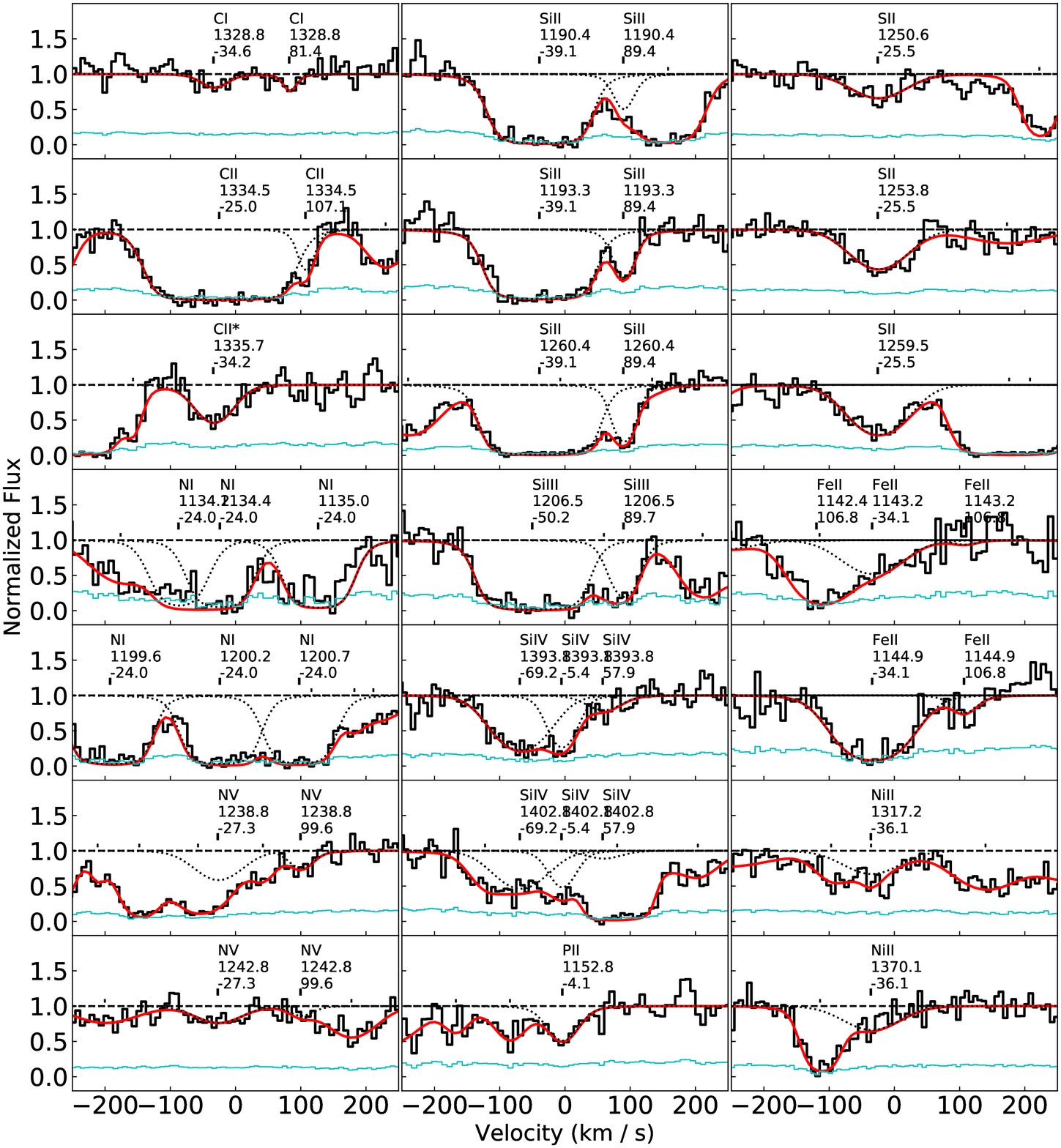}
\end{center}
\caption{The spectra and best-fit models for identified lines for NGC 891. The black histograms are the observed spectrum, while the cyan lines are the error. The red solid lines are the total model for all identified lines, while the black dotted lines are decomposed spectral lines associated with NGC 891. The label for each identified line is the ion, the wavelength and the velocity relative to $z = 0.001761$.}
\label{lines}
\end{figure*}

To coadd different exposures, we follow the procedure described in \citet{Wakker:2015aa}, for which we give a brief summary here.
In each exposure, we calculate the net gross counts using the net rate and the exposure time, which is corrected for the background and the fixed pattern noise.
Then, we coadd the total counts from each exposure and divide by the total exposure time at each wavelength to obtain the net count rate.
The final noise is the Poisson noise derived from the total counts, which is found to be better matched with the measured noise of the coadded spectrum \cite{Wakker:2015aa}.
It is known that COS spectra may have velocity shifts ($\approx 20 \kms$) due to the geometric distortion and wavelength solutions from CALCOS.
However, in this spectrum, the continuum count rates are too low (most pixels have zero/one gross count in one exposure) to do the cross correlation or to measure line centroids of single lines, which is necessary to correct the wavelength solution of individual exposures.
Therefore, we adopt the CALCOS wavelength solutions and do not align different exposures using strong Galactic lines in this stage.
The median $S/N$ is 7.5 in the coadded COS/FUV spectrum at $R=20000$ and 14.4 in the coadded COS/NUV spectrum at $R=2500$.

The coadded spectrum is binned by three pixels (yielding the bin width of $\Delta\lambda = 0.02991\rm~\AA$) for fitting and line identification in NGC 891 (see the Appendix for details for other systems).
Twelve ions are detected near the systemic redshift of $528 \kms$ (z = 0.001761; \citealt{de-Vaucouleurs:1991aa}): \ion{C}{1} ($2.1\sigma$), \ion{C}{2}, \ion{C}{2}$^*$, \ion{Fe}{2}, \ion{Si}{2}, \ion{Si}{3}, \ion{Si}{4}, \ion{S}{2}, \ion{N}{1}, \ion{N}{5}, \ion{Ni}{2}, and \ion{P}{2}.
In addition to the primary absorption system, there is a weaker absorption system at $v = 90-100 \kms$ related to the systemic velocity of $528 \kms$, which we refer as a high velocity cloud (HVC).
The properties measured from fitting Voigt profiles to these lines are given in Table \ref{NGC891}, and the best-fit models are shown in Fig. \ref{lines}.
For each ion, different lines are fitted simultaneously and the continuum is determined using spline fitting in intervals of $3-6\rm~\AA$.
The Voigt profile model is convolved with the COS line spread function to address the non-Gaussian broadening of COS, and the preferred solution is determined by minimizing the $\chi^2$. 
Extra one component is added if the $\chi^2$ difference is larger than 10 ($> 2 \sigma$ at degrees of freedom of 3).

\begin{figure*}
\begin{center}
\includegraphics[width=0.96\textwidth]{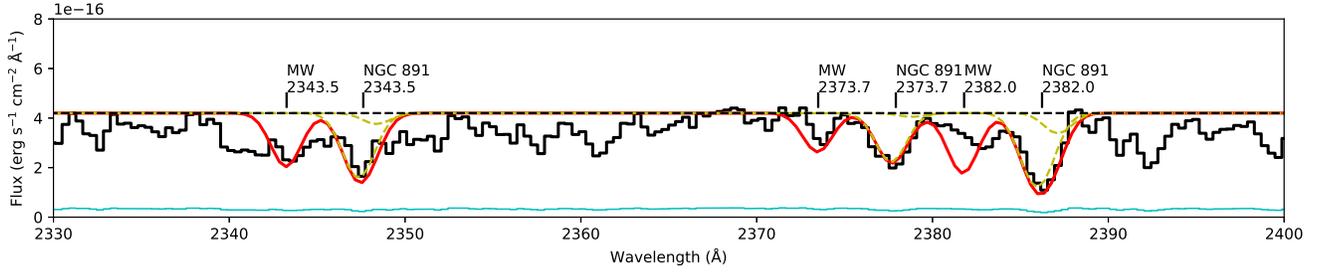}
\end{center}
\caption{The \ion{Fe}{2} lines from the FUV spectrum are projected onto the NUV spectrum. The continuum level is set to be a constant at $4.2\times 10^{-16}\rm~erg~s^{-1}~cm^{-2}~\AA^{-1}$. The solid red line is the projected \ion{Fe}{2} using the measurements from the FUV band, while the dashed yellow lines are the major absorption component and the HVC of NGC 891. The NGC 891 \ion{Fe}{2} line strengths are consistent between the FUV and the NUV spectra, while the MW \ion{Fe}{2} may be affected by fixed pattern noises.}
\label{FeII}
\end{figure*}

In the primary absorption features at $\approx 0\kms$, \ion{N}{1}, \ion{Fe}{2}, \ion{C}{2}, \ion{Si}{2}, and \ion{Si}{3} show saturated features, which might affect the column density measurements.
For saturated lines, the column densities are degenerate with $b$ factors that cannot be well determined using single lines.
For \ion{N}{1}, \ion{Fe}{2}, and \ion{Si}{2}, multiple lines from the same ion break the degeneracy between the $b$ factor and the column density.
\ion{N}{1} and \ion{Fe}{2} have multiple lines including unsaturated lines, which constrains the column densities and $b$ factors.
The situation is different for \ion{Si}{2}, since all of the three observed lines are saturated.
The simultaneous fitting of multiple lines leads to acceptable measurement uncertainties ($\log N = 15.60\pm0.26$ and $b=39.2\pm2.9$).
However, this measurement may have additional systematic errors that are dependent on the coadding method and the continuum determination, which is more sensitive for saturated lines.
Therefore, besides the direct measurement from the Voigt profile fitting, we also determine a lower limit of $\log N >14.8$ for \ion{Si}{2} at the $1\sigma$ level.
For \ion{Si}{3}, there is only one absorption line, which would result in the lower limit of the column density.
One possible solution is using the \ion{Si}{2} $b$ factor to constrain the $b$ of \ion{Si}{3}.
As adjacent ions, \ion{Si}{2} and \ion{Si}{3} are believed to be in a similar phase, which leads to similar $b$ factors. 
We adopted the column density by fixed \ion{Si}{3} $b$ factor to $40 \kms$ (the \ion{Si}{2} $b$ factor), leading to a column density of $\log N =15.11\pm0.24$.
Otherwise, we obtain the lower limit for \ion{Si}{3} similar to \ion{Si}{2}, and the $1\sigma$ lower limit is $\log N = 14.7$ for \ion{Si}{3}.
Therefore, we have two sets of measurements for \ion{Si}{2} and \ion{Si}{3} (Voigt profile fittings and lower limits), which will be used in the modelings (Section 3.1).

For \ion{C}{2}, we cannot assume that \ion{C}{2} has the same $b$ factor as \ion{C}{1} (low significance) or \ion{C}{2}*, since a \ion{C}{2} $b$ factor of $23-33\rm~ km~s^{-1}$ leads to an unrealistically large column density ($\log N >18$) and wide line wings that are not observed.
This might be because \ion{C}{2}* is from a dense region with a small $b$, while the \ion{C}{2} line probably includes additional components from low-density ionized gases (e.g., the HVC at $v=-88\kms$). 
However, the line shape of \ion{C}{2} cannot be decomposed because it is saturated.
Therefore, we varied the $b$ factor to give a lower limit of the column density.
As a constraint, the $b$ factor cannot be lower than $35\kms$ or higher than $66\kms$ ($1 \sigma$ bounds), with $\log N=18.5$ to 15.8, correspondingly.
In the following analysis (Section 3.1), we do not include \ion{C}{2} to constrain the photoionization model, but we will consider the consistency of \ion{C}{2} with other ions.

In the HVC, we detected absorption lines from \ion{C}{2}, \ion{Si}{2}, \ion{Si}{3}, \ion{N}{5}, and \ion{Fe}{2}.
\ion{C}{1} is detected at the significance level of $1.7\sigma$ with a velocity separation of $81.4 \pm5.7 \kms $, which is more than $2\sigma$ different from most other ions. 
\ion{Si}{3} is contaminated by the \ion{Ne}{6} lines at $z=1.1655$ (an AGN outflow with multiple components over $1000\kms$;  see the Appendix for details). 
A possible $v=-170\kms$ at $z=1.1655$ component of the AGN outflow (shown by \ion{O}{5}, \ion{Ne}{5}, and \ion{Ar}{7}) is near the  $v=86\kms$ component of the NGC 891 \ion{Si}{3} line $\lambda 1206.5\rm ~\AA$.
This AGN outflow component affects the line shape of the HVC \ion{Si}{3} line, which may lead to a smaller $v$ value. 
Both \ion{Si}{3} and \ion{Ne}{6} are isolated in the wavelength coverage of the current QSO spectrum, so one cannot break this degeneracy using other lines from these two ions.
To remove this degeneracy, we fixed the velocity of \ion{Si}{3} to that of \ion{Si}{2} (see Appendix) .
The $b$ factor of this \ion{Si}{3} cannot be well constrained, but it should lie between $15\kms$ (\ion{Si}{2}) to $20\kms$ (\ion{N}{5}).
The adopted column density is $\log N = 13.37\pm0.14$ for the \ion{Si}{3} HVC component with $b=20\kms$ and $\log N=13.68\pm0.25$ with $b=15\kms$.
The difference between these two values has complex reasons: the possible saturation of this line and the uncertainty of the AGN outflow \ion{Ne}{6} line.
Therefore, we set a lower limit of $\log N=13.1$ for the HVC \ion{Si}{3}, and the upper limit of this ion has a large uncertainty.
In Section 3.2, we will show that it is the lower limit of \ion{Si}{3} that constrains the photoionization model, so the uncertainty of the upper limit of the HVC \ion{Si}{3} will not affect our photoionization modeling (Fig. \ref{hvc_model}).
\ion{Si}{4} has a component at $v=57.9\pm 21.8 \kms$, which may or may not be associated with the NGC 891 HVC.
In following analysis we assume it is associated with the major component at $v=-30 \kms$.

For the HVC, we also measure the upper limit of the \ion{N}{1} and \ion{Si}{4} column densities, which may have constraints on the physical conditions. 
The 2$\sigma$ upper limits for \ion{Si}{4} ($\lambda 1393.8$\AA) and \ion{N}{1} ($\lambda$1200.7\AA) are $\log N < 12.5$ and 13.8 cm$^{-2}$, respectively.

The COS/NUV G230L spectrum has a lower spectral resolution, so the line measurements (e.g., $b$ factors) are less accurate.
There is only one detectable ion (i.e., \ion{Fe}{2}) in this band, which is also detected in the FUV band.
Therefore, we only use this NUV spectrum to check the consistency of \ion{Fe}{2} measurements, and we project the  \ion{Fe}{2} properties obtained from the FUV spectrum into the NUV lines (Fig. \ref{FeII}).
The NUV spectrum is consistent with the measurements obtained from the FUV spectrum, although the MW \ion{Fe}{2} shows some differences. 
We suggest that the NUV MW \ion{Fe}{2} might be affected by unknown fixed pattern noise.
However, the individual exposures have too few counts to check for this effect.
The measurements from the FUV spectrum are adopted in the following analysis for three reasons.
First, the FUV spectrum shows two matched strong lines with significance $>6\sigma$, which is unlikely to be by chance.
Second, four of the NUV lines from both the MW and NGC 891 are consistent with the FUV measurements.
Third, the COS/NUV G230L spectrum is different from the STIS spectrum at the same wavelengths.
The STIS spectrum shows stronger absorption around $2382.0\rm~\AA$ \citep{Bregman:2013aa}, which corresponds to the unmatched MW line \ion{Fe}{2} $\lambda 2382.0 \rm~\AA$ in the COS/NUV spectrum.

\begin{table}
\tablewidth{1.0\columnwidth}
\begin{center}
\caption{\ion{H}{1} 21 cm Line Fitting Results}
\label{HI21_fit}
\begin{tabular}{cccccc}
\hline
\hline
$\log N$ & $\sigma_N$ & $b$ & $\sigma_b$ & $v$ & $\sigma_v$ \\
\hline
\multicolumn{6}{c}{MW (LQAC 035+042 003)} \\
\hline
20.32 & 0.01 & 27.49 & 0.62 & -24.50 & 0.87 \\
20.35 & 0.01 & 13.11 & 0.23 & -3.34 & 0.15 \\
19.24 & 0.01 & 1.29 & 0.05 & 0.53 & 0.02 \\
20.22 & 0.01 & 5.71 & 0.08 & 1.64 & 0.06 \\
19.16 & 0.02 & 1.52 & 0.08 & 5.15 & 0.04 \\
\hline
\multicolumn{6}{c}{NGC 891} \\
\hline
19.05 & 0.10 & 10.9 & 2.4& -88.7 & 1.6 \\
19.82 & 0.06 & 30.5 & 4.1 & -27.4 & 3.0 \\
19.58 & 0.10 & 20.5 & 2.9 & 19.0 & 2.6 \\
19.08 & 0.09 & 12.6 & 2.6 & 114.7 & 1.9 \\
\hline
\multicolumn{6}{c}{MW (3C 66A)} \\
\hline
19.65 & 0.13 & 40.93 & 4.30 & -37.39 & 4.66 \\
20.13 & 0.02 & 17.20 & 0.76 & -34.64 & 0.42 \\
19.70 & 0.02 & 6.13 & 0.21 & -19.53 & 0.12 \\
20.58 & 0.01 & 12.02 & 0.12 & -2.26 & 0.08 \\
19.56 & 0.03 & 2.34 & 0.11 & -0.25 & 0.16 \\
20.14 & 0.01 & 2.56 & 0.04 & 3.07 & 0.06 \\
\hline
\end{tabular}
\end{center}
\end{table}

\begin{figure*}
\begin{center}
\includegraphics[width=0.48\textwidth]{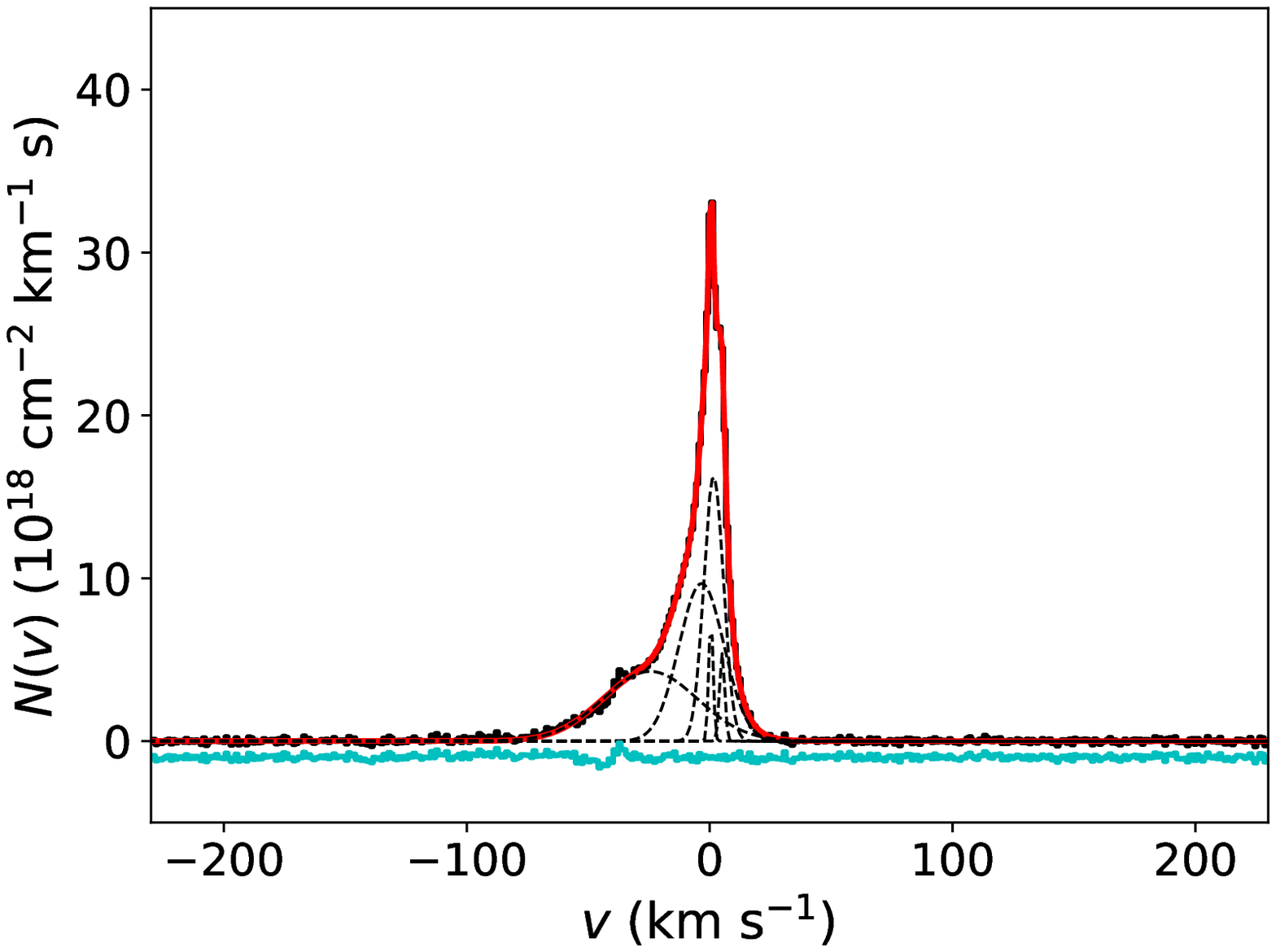}
\includegraphics[width=0.48\textwidth]{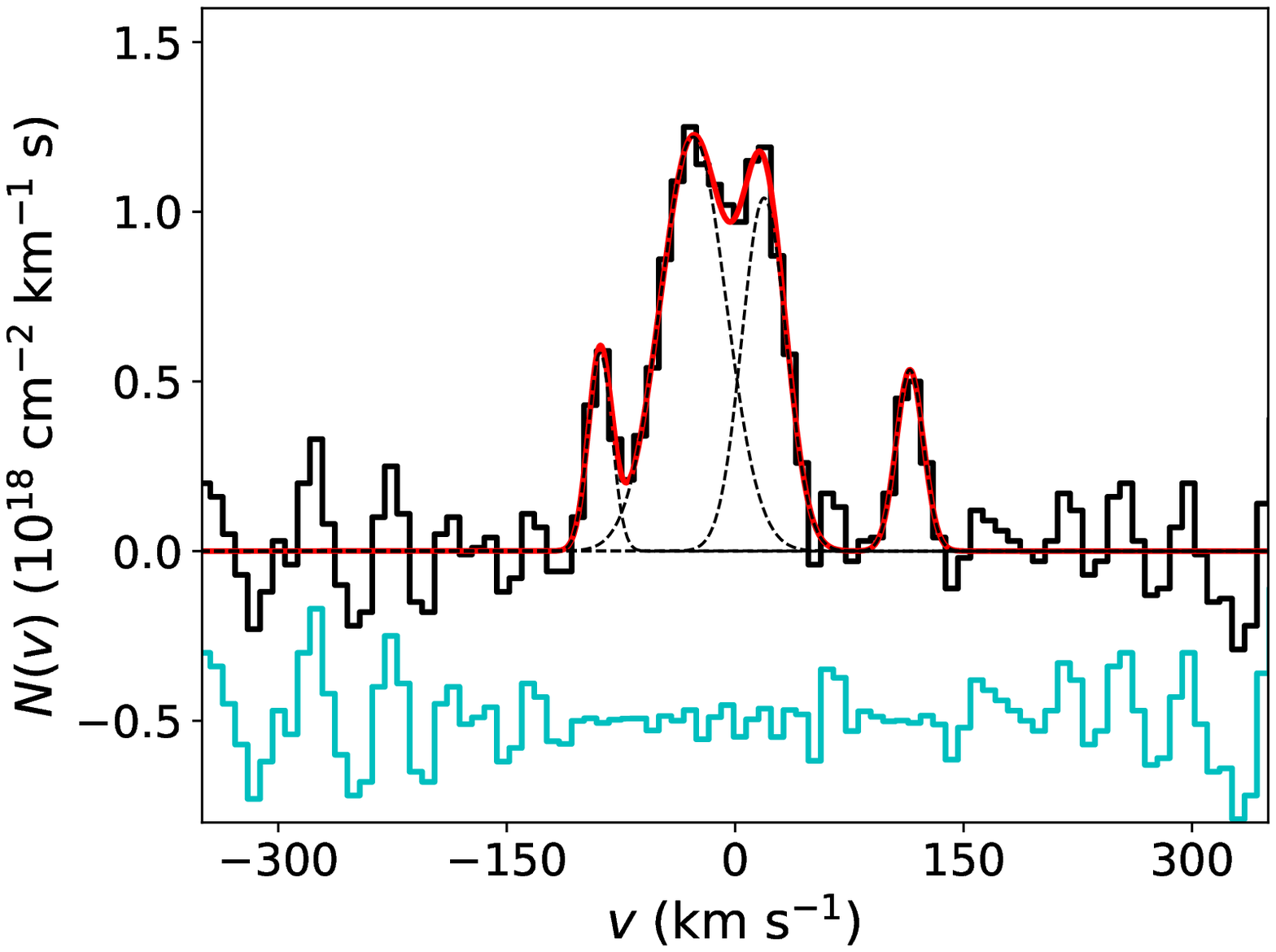}
\end{center}
\caption{The \ion{H}{1} 21 cm line fitting results for both the MW ({\it left panel}) and NGC 891 ({\it right panel}). Each line is fitted by five (MW) / four (NGC 891) Gaussian components and results are shown in Table \ref{HI21_fit}. The data and the fit are shown in the black and red solid lines respectively, while individual components are shown in the black dashed lines. The cyan lines are the residuals shifted by $-1\times 10^{18} \rm ~cm^{-2} ~km^{-1} ~s$ for the MW and $-5 \times 10^{17} \rm ~cm^{-2} ~km^{-1} ~s$ for NGC 891.}
\label{HI21}
\end{figure*}

\begin{figure*}
\begin{center}
\includegraphics[width=0.96\textwidth]{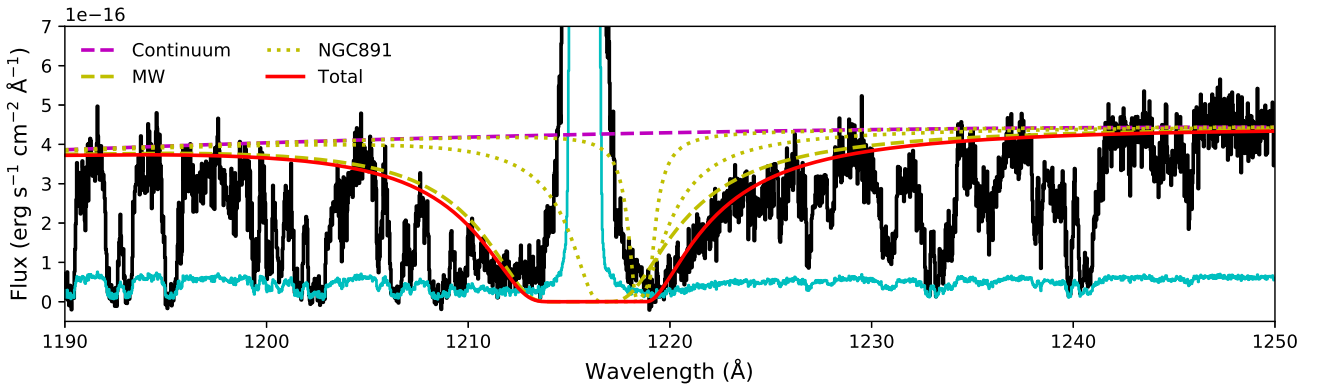}
\end{center}
\caption{The projected Ly$\alpha$ absorption feature. The MW component (the dashed yellow line) is the summation of four Gaussian components fitted to the \ion{H}{1} 21 cm line. The NGC 891 contribution is divided into the two components: the major absorption at $v=-30 \kms$ (the broad component in the dotted yellow lines); and the HVC at $+100\kms$ (the narrow component). The continuum (the magenta dashed line) is a second order polynomial function. The total model (the red solid line) matches the wing feature of the observed Ly$\alpha$.}
\label{Lya}
\end{figure*}

\subsubsection{The STIS Spectrum}
The STIS spectrum is reduced and reported in \citet{Bregman:2013aa}; therefore, we do not describe the STIS observational details here.

From the STIS spectrum, \citet{Bregman:2013aa} obtained the \ion{Fe}{2} column density of $\log N = 14.44\pm 0.14$,  the \ion{Mg}{1} upper limit of $12.67$ ($\rm EW = 0.50\pm0.17\rm~\AA$), and \ion{Mg}{2} of $15.48^{+0.36}_{-0.27}$.
\ion{Mg}{1} and \ion{Mg}{2} can not be measured in the COS data, but the \ion{Fe}{2} measurement from COS is $\log N = 15.10\pm 0.04$.
The \ion{Fe}{2} in the STIS spectrum is measured from the line $\lambda 2343.5\rm~\AA$, since two other lines ($\lambda 2373.7\rm~\AA$ and $\lambda 2382.0\rm~\AA$) are completely blended with the Galactic \ion{Fe}{2} lines.
However, the line $\lambda 2343.5\rm~\AA$ is also partially blended with the Galactic line.
Therefore, we adopt the measurements from the COS/FUV spectrum, which is supported by the matched and separated COS/FUV and COS/NUV spectral lines (Fig. \ref{lines} and Fig. \ref{FeII}).

We update the measurements of \ion{Mg}{1} with accurate dynamic information from the high resolution FUV spectrum.
\ion{Mg}{1} is an isolated line in STIS spectrum with a velocity of $v= -43 \pm 79 \kms$, which is consistent with lines from \ion{N}{1}, \ion{C}{1}, and \ion{C}{2}*.
These ions should be cospatial, considering the major contributor of  \ion{C}{1}, \ion{N}{1}, and \ion{Mg}{1} is the neutral phase, and \ion{C}{2}* traces the high density gas that is also more neutral than other ions (e.g., \ion{C}{2} or \ion{Fe}{2}).
Therefore, we assume the \ion{Mg}{1} also has a $b$ factor of $33 \kms$ (the thermal broadening is ignored, which is less than $10 \kms$), which is the $b$ factor from \ion{C}{2}* and \ion{N}{1}.
The \ion{C}{1} $b$ factor is not preferred because \ion{C}{1} is at a low significance ($<3 \sigma$) leading to a large uncertainty.
Then, the \ion{Mg}{1} column density is determined to be $\log N = 12.76 \pm 0.24$.

The Doppler $b$ factor of \ion{Mg}{2} should be similar to \ion{Fe}{2}, \ion{Ni}{2}, etc., which have $b$ factors around $55\kms$.
This $b$ factor is consistent with the previously assumed value of $56\kms$ from the width of the \ion{H}{1} 21 cm line \citep{Bregman:2013aa}.
However, \citet{Bregman:2013aa} determined the \ion{Mg}{2} column density using only the line at \ion{Mg}{2} $\lambda 2802.7\rm~\AA$ ($\log N = 15.48_{-0.27}^{+0.36}$). 
For the \ion{Mg}{2} $\lambda 2795.5.\rm~\AA$ line, we measured $\log N = 15.3_{-0.9}^{+1.2}$ ($\rm EW = 2.31 \pm 0.61 ~\AA$ assuming $b=55\kms$).
There are two undetectable \ion{Mg}{2} lines $\lambda 1239.9\rm ~\AA$ and $\lambda 1240.4\rm~\AA$ in COS/FUV spectrum, so we also obtain constraints from these two lines.
Using the limiting EW of $12.0\rm~m\AA$ ($1\sigma$), the upper limit of the \ion{Mg}{2} column density is $\log N = 15.5$ ($2\sigma$) or  $\log N = 15.1$ ($1\sigma$).
Therefore, we prefer the measurement from the \ion{Mg}{2} $\lambda 2795.5.\rm~\AA$ line, and adopt $\log N = 15.3\pm 0.3$ for the \ion{Mg}{2} column density.

\subsubsection{The \ion{H}{1} 21 cm Line}
The MW \ion{H}{1} 21 cm line is obtained from the Leiden/Argentine/Bonn (LAB) Survey of Galactic \ion{H}{1} with an effective beam size of $0.2^\circ$ around the QSO \citep{Kalberla:2005aa}. 
The \ion{H}{1} 21 cm line data of NGC 891 is from \citet{Oosterloo:2007aa}, and extracted around the QSO with a beam size of $< 1.0'$. 
Both \ion{H}{1} 21 cm lines show multiple components, and we fit the spectra using Gaussian functions, shown in Fig. \ref{HI21} and Table \ref{HI21_fit}.

For NGC 891, there are two high velocity clouds shown in the \ion{H}{1} 21 cm line: $-88 \kms$ and $+114\kms$. 
The component at $+114 \kms$ is observed in multiple ions, while the HVC at $-88\kms$ is only probably detected in \ion{Si}{4} with a velocity difference of $\approx 20 \kms$ ($> 2\sigma$).
Therefore, we do not consider this component as a separate HVC in the following analysis.
The NGC 891 absorption system is divided into two components -- the major component at $v= -30\kms$ and the HVC at $v=+100\kms$. 
The summation of the three low-velocity components leads to a total \ion{H}{1} column density of $\log N = 20.06\pm 0.08$.
For the HVC at $110\kms$, the \ion{H}{1} column density is $\log N = 19.08\pm 0.09$.

\begin{figure*}
\begin{center}
\includegraphics[width=0.96\textwidth]{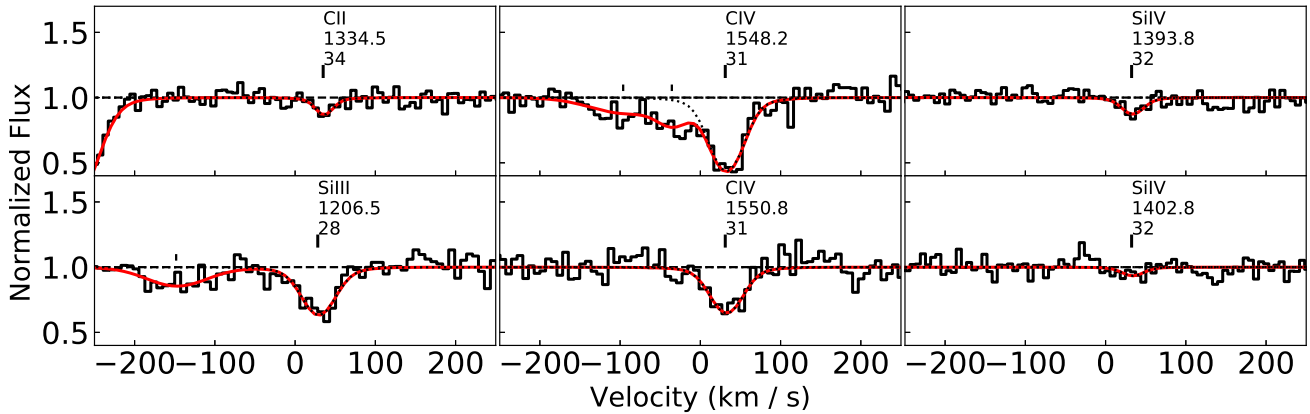}
\end{center}
\caption{The spectra and best-fit models for identified lines toward 3C 66A and near the systemic velocity of NGC 891. The red solid lines are the total model for all identified lines, while the black solid lines are spectral lines associated with NGC 891.}
\label{3c_lines}
\end{figure*}

We examine whether the 21 cm line is consistent with the FUV Ly$\alpha$ absorption feature to exclude a possible contamination within the radio beam.
The FUV spectrum should be aligned with the \ion{H}{1} 21 cm line spectrum, since the COS/FUV spectrum may have a velocity shift up to $\approx 30\kms$ \citep{Wakker:2015aa}.
We employ galactic \ion{Si}{3}, \ion{C}{1}, and \ion{C}{2}* lines to do this alignment, since \ion{N}{1} may be affected by geocoronal emission.
The mean velocity in the FUV spectrum is $-16.3 \kms$ ($13.7 \kms$ at $z = -0.0001$), while the peak of the \ion{H}{1} 21 cm line is approximately $1.6 \kms$.
Therefore, the velocity difference is $17.9\kms$ and the corrected velocities are reported in Table \ref{NGC891}.
Here, the LAB \ion{H}{1} 21 cm lines are in the local standard of rest frame, while other spectra are in the heliocentric frame.
The difference between these two frames is a constant ($1.4\kms$) over different wavelengths.
Applying the velocity shift of $17.9\kms$, we correct the COS/FUV wavelength calibration, and convert the COS/FUV spectrum into the local standard of rest frame.

In Fig. \ref{Lya}, we project the fitted \ion{H}{1} from 21 cm lines into the Ly$\alpha$ absorption including the velocity shift.
The two components of NGC 891 are plotted separately and the QSO continuum is fitted by a second order polynomial function within $1190-1250\rm~\AA$.
Against the Ly$\alpha$ wing, other absorption features occurs, so we compare the projected Ly$\alpha$ shape with the continua at $\lesssim 1208\rm~\AA$ and $\gtrsim 1220\rm~\AA$.
The red wing of the Ly$\alpha$ ($\lesssim 1224\rm~\AA$) matches with the \ion{H}{1} 21 cm line projection, while in the blue wing ($\lesssim 1208\rm~\AA$), the continuum region around $1205\rm~\AA$ shows consistency between the UV and radio observations.
Therefore, we suggest that the average \ion{H}{1} in the radio beam is consistent with the FUV Ly$\alpha$ absorption lines.

\subsection{3C 66A}
\subsubsection{The AGN}
3C 66A is a BL Lac AGN at $\rm RA = 02h22m39.6s $ and $\rm DEC = +43^\circ 02' 08''$ with the redshift of $z=0.444$, which is $41.2'$ away from the galactic center of NGC 891. This AGN is close to the major axis of NGC 891, which is offset from the position angle of the major axis by $21.8^\circ$ (clockwise) north of the disk and is in the same side as the QSO LQAC 035+042 003. The impact parameter of this sightline is  $108.3\rm~kpc$ and it is $18.4\rm~kpc$ above the disk mid-plane.

\subsubsection{The {\it HST}/COS Spectrum}
The COS spectra of 3C 66A were observed in the {\it HST} program 12612 (1 Nov 2012; PI: Stocke) and 12863 (8 Nov 2012; PI: Furniss). 
The total exposure times are $12.6\rm~ks$ for the G130M and $7.2\rm~ks$ for the G160M. 
Using the data reduction processes similar to LQAC 035+042 003, we coadd the COS spectra for 3C 66A. 
The coadded spectra have median $S/N$ values of $25$ for the G130M and $15$ for the G160M.

The line list is available in \citet{Danforth:2016aa}; therefore, we focus on the absorption system associated with NGC 891.
The measured line properties are listed in Table \ref{NGC891}, while the best-fit models are shown in Fig. \ref{3c_lines}.
Our measurements are all consistent with \citet{Danforth:2016aa}, except for the detectable \ion{C}{2} in our spectrum with a significance of $3.4\sigma$. 
For these weak features, it is possible that the coadd process could affect the completeness.

We also measure the upper limit for \ion{Si}{2} of $\log N < 11.90$ at $2 \sigma$, which is obtained from the line \ion{Si}{2} $\lambda 1260.4\rm~\AA$ assuming $b=20\kms$ (similar to \ion{Si}{3}).

\subsubsection{The \ion{H}{1} 21 cm Line}
The MW \ion{H}{1} 21 cm line spectrum is also from the LAB survey. 
Following the previous steps, we fit the \ion{H}{1} 21 cm line and project the shape into the Ly$\alpha$ feature.
The fitting results are shown in Table \ref{HI21_fit}.
The projected Ly$\alpha$ agrees with the FUV spectrum, and there are no obvious components due to the NGC 891 absorption features, which gives an upper limit of $\log N ({\rm HI}) < 19.5$ at the $2 \sigma$ level.

Again, we obtain a velocity of $-19.7 \kms$ at $z=0.0$ for the 3C 66A FUV spectrum using the Galactic \ion{N}{1} and \ion{O}{1} lines, and a shift of $22.8\kms$ between the FUV spectrum and the \ion{H}{1} 21 cm line.

\section{Model}
Based on the measured ion column densities, we estimate the ionization parameter, the gas temperature, and the metallicity in the ionization equilibrium model with the photoionization modification.
The density of the absorption system with $v=-30\kms$ can be derived separately from the density sensitive line \ion{C}{2}* $\lambda 1335.7\rm~\AA$.

\begin{figure*}
\begin{center}
\includegraphics[width=0.48\textwidth]{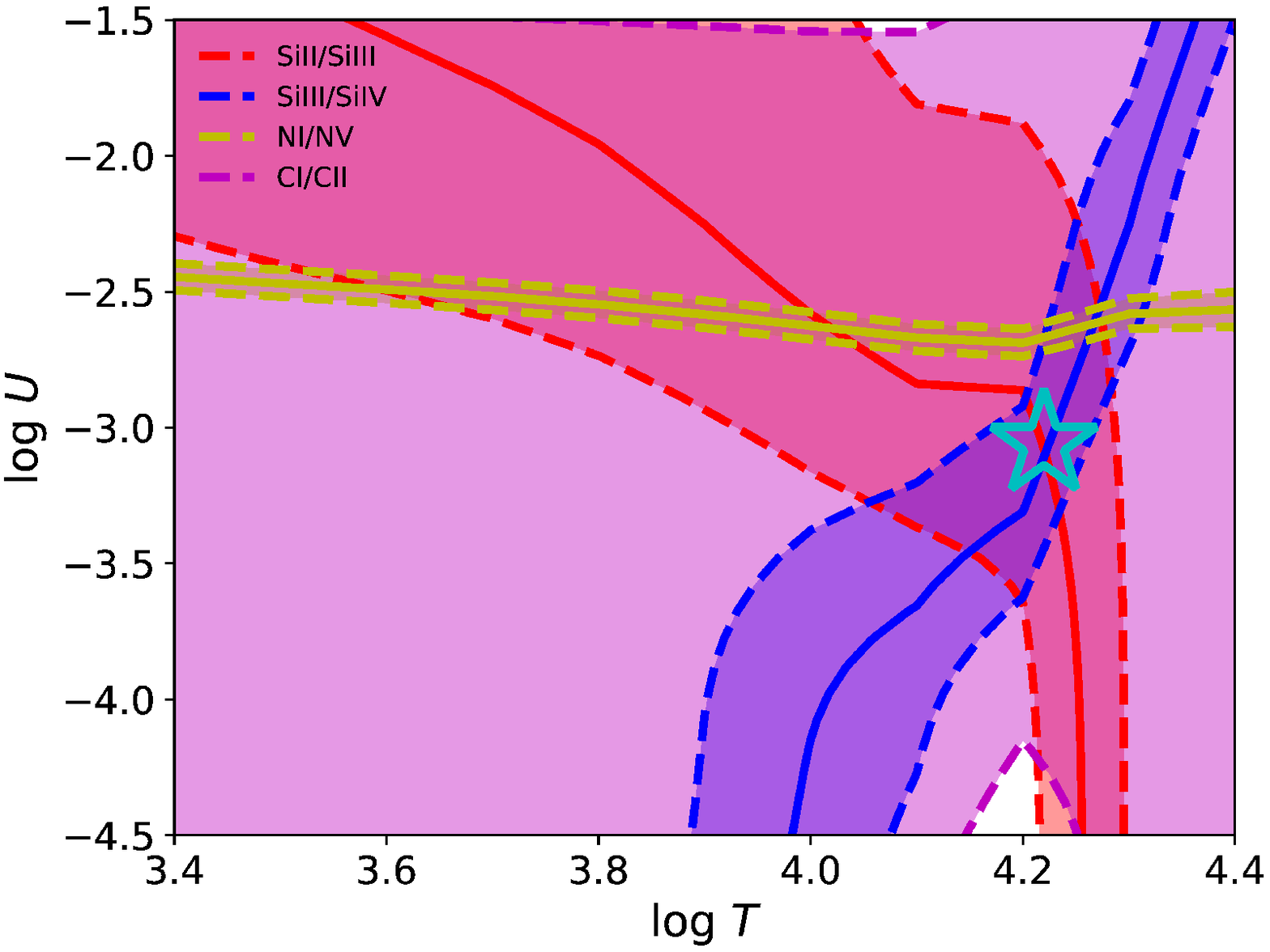}
\includegraphics[width=0.48\textwidth]{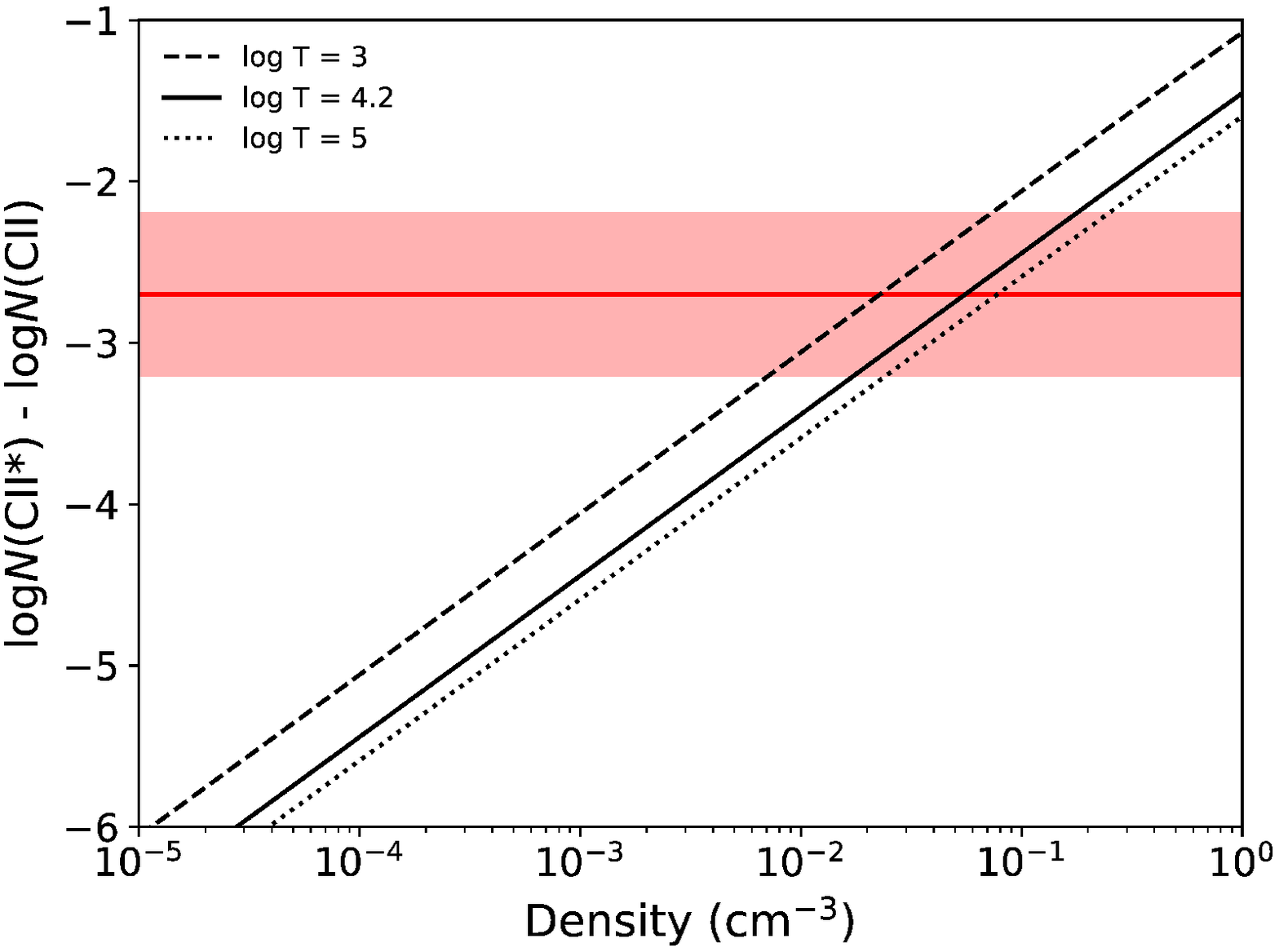}
\end{center}
\caption{{\it Left panel:} the photoionization model of the $v=-30\kms$ system. The shadowed belts are the acceptable regions ($1\sigma$) for each ion column density ratio. The open cyan star is the preferred solution of $\log U = -3.06\pm0.10$ and $\log T = 4.22\pm 0.04$. {\it Right panel:} the density measurement from the \ion{C}{2}*/\ion{C}{2} ratio. The colored bar is the $1\sigma$ uncertainty of the observed \ion{C}{2}*/\ion{C}{2} ratio, while the black lines are the CHIANTI predictions at different $\log T$ of 3.0 (dashed), 4.2 (solid) and 5 (dotted).}
\label{model_major}
\end{figure*}

\begin{figure}
\begin{center}
\includegraphics[width=0.48\textwidth]{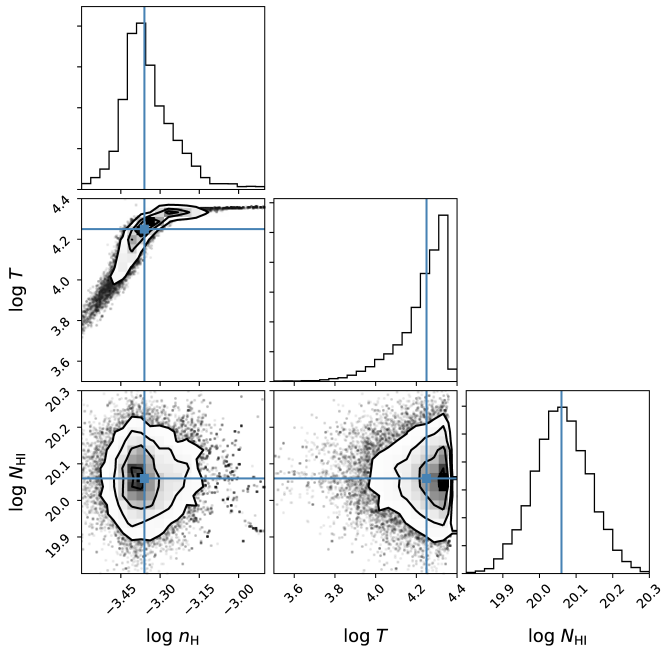}
\end{center}
\caption{The marginalized posterior distributions in MCMC model for the major absorption system at $v\approx 30\kms$. The cyan lines indicate the medians for all parameters.}
\label{model_major_mcmc}
\end{figure}

\subsection{The Low Velocity System at $v=-30\kms$ toward LQAC 035+042 003}
\label{major_absorption}
\subsubsection{The Photoionization Model}
To build the photoionization model, we employ the photoionization code Cloudy (version 17.00; \citealt{Ferland:2013aa}).
Due to the proximity to the galactic center of NGC 891, the radiation originating from the galaxy disk cannot be ignored during the calculation.
Therefore, the incident field should be the summation of the universal ultraviolet background (UVB) and the escaping light from the disk. 
However, the escaping light is difficult to determine and there is no published radiation field of NGC 891.
We assume the escaping light has the same shape of the UVB at low energies, where the observed low ionization state ions of NGC 891 occur.
Then, although the incident field is more intense than the UVB, we can use the UVB-only field to determine the ionization parameter.
In this case, we adopt the UVB at $z=0.0$ from \citet{Haardt:2012aa} in the following calculation to set the shape of the incident field.
The uncertainty of this assumption is discussed in Section 3.1.3.

In the Cloudy model, we fix the neutral hydrogen column density to the measured value from the \ion{H}{1} 21 cm line (i.e., $\log N = 20.06$), and the metallicity is fixed to the solar metallicity \citep{Asplund:2009aa}.
We explore the parameter space for the density of $\log n_{\rm H} = -5.0$ to $-1.5$ (roughly ionization parameter $\log U = -1.5$ to $-5.0$) and the temperature of $\log T = 3.4$ to $4.4$.

Using the different ionization state ions from one element, one can obtain the constraints on the ionization parameter and the temperature without involving element abundances.
As shown in Fig. \ref{model_major}, we consider three elements: silicon, carbon and nitrogen, which have two or more observed ions.
Silicon has three consecutive ions, which could give a best solution for the gas phase properties.
Adopting measurements from the Voigt profile fitting, we obtain $\log U = -3.06\pm0.10$ ($\log n_{\rm H} = -3.44 \pm 0.10$) and $\log T = 4.22 \pm 0.04$.
This solution has a weak dependence on the metallicity; therefore, we use the solar metallicity in our model \citep{Asplund:2009aa}, since NGC 891 is similar to the MW.

Besides this solution, we also build a Markov chain Monte Carlo (MCMC) model to constrain the properties using the lower limits of \ion{Si}{2} and \ion{Si}{3}.
In this model, we follow the method in \citet{Fumagalli:2016aa}, adopting the Gaussian function as the likelihood for good measurements (i.e., \ion{Si}{4}), and a rescaled cumulative distribution function for lower limits (i.e., \ion{Si}{2} and \ion{Si}{3}).
We employed {\it emcee} to sample the parameter space with 100 walkers and 500 steps, and the first 50 steps are masked out as the thermalization stage \citep{Foreman-Mackey:2013aa}.
The final solution is similar to the solution using Voigt fitting results, but with larger uncertainties.
As shown in Fig. \ref{model_major_mcmc}, the gas properties are $\log U = -3.14^{+0.07}_{-0.12}$ ($\log n_{\rm H} = -3.36_{-0.07}^{+0.12}$), $\log T = 4.25_{-0.15}^{+0.08}$, and $\log N_{\rm H} = 20.06_{-0.08}^{+0.08}$.
There are two reasons for the larger uncertainty: the inclusion of the \ion{H}{1} uncertainty in the model; and the lower limits of \ion{Si}{2} and \ion{Si}{3}.
In the ratio-matched model, the \ion{H}{1} column density is fixed as a constant, while in the MCMC model, the variation of \ion{H}{1} column density could lead to a larger uncertainty.
Compared to the measurements with two-sided constraints, only using the lower limits softens the constraints.
The reason for the similarity between the two models is because the acceptable parameter space is determined by the lower limit of the \ion{Si}{3} column density and the \ion{Si}{4} column density.
The measurements of \ion{Si}{4} set a strong constraint of the temperature of $\log T<4.4$ as shown by the sharp turnover in Fig. \ref{model_major_mcmc}, which is due to the ionization fraction peak of \ion{Si}{4} (the parameter space can extend to $\log T = 5$).
The lower limit of \ion{Si}{3} leads to a lower limit of the temperature; therefore, the temperature can be determined well.
In the MCMC solution, the predicted column densities of these ions has a systematic difference compared to the ratio-matched model.
The MCMC model predicts higher column densities by $0.1-0.3$ dex for 11/13 metal ions, except for two relatively high ionization state ions (\ion{Si}{4} and \ion{N}{5}).
This systematic difference occurs because this model predicts a higher \ion{H}{2} column density ($\log N = 21.01$ compared to $20.68$).
With a difference of $0.3$ dex, both models predict that the major absorption system is dominated by the ionized gases, which is about $5-10$ times more massive than the \ion{H}{1} gases.

The ratio between \ion{C}{1} and \ion{C}{2} has a large uncertainty due to the uncertain \ion{C}{2} measurement.
Therefore, we only check the consistency between the \ion{C}{2} line and the model prediction.
Applying the \ion{C}{1} to \ion{C}{2} ratio of the preferred solution from the silicon ions, the predicted \ion{C}{2} column density is $\log N=16.8$.
Direct Voigt profile fitting to the \ion{C}{2} line leads to $\log N=16.4 \pm 0.5$ and $b=53\pm8\kms$ with a total $\chi^2=292.6$ and a degree of freedom (dof) of $256$.
By fixing the $b$ factor at $47\kms$, the corresponding column density is $\log N=16.8$ with $\chi^2=293.1$, which only increases the $\chi^2$ by 0.5.
Therefore, the predictions from the preferred model are consistent with the \ion{C}{2} observation.

The ratio between \ion{N}{1} and \ion{N}{5} is not consistent the preferred solution, as the \ion{N}{5} column density is too high, which indicates that \ion{N}{5} is not in the same phase as low ionization state ions.
The \ion{N}{4} to \ion{N}{5} ionization potential, 77.5 eV, is twice the ionization potential from \ion{Si}{3} to \ion{Si}{4} ($33.5\rm~ eV$; the second highest potential among detected ions).
Therefore, we suggest that \ion{N}{5} is not produced in the same phase of the low ionization state gas.

In the preferred model ($\log U = -3.06$ and $\log T = 4.22$), the total hydrogen column density $\log N = 20.81$. 
By comparing the model column densities with the observed ion column densities, we calculate the residuals for different ions, which indicates the element abundances.
The results are summarized in the Table \ref{metallicity}.
For each element, the relative abundance is calculated as the average value for all ions belonging to the element.
The uncertainty of the abundance is affected by two factors -- the uncertainty of the photoionization model (assumed to be $\sigma_{\rm PI} = 0.2$ dex) and the observed uncertainty of the total column densities for one element ($\sigma_{N}$).
The total uncertainty is $\sigma_{\rm [X/H]} = (\sigma_{\rm PI}^2 + \sigma_N^2)^{1/2}$.
For magnesium, the uncertainty is the scatter between the measurement of \ion{Mg}{1} and \ion{Mg}{2}.
As stated previously, we noticed that there is a systematic difference for the column densities of metal ions between the ratio-matched model and the MCMC model, which is mainly due to the higher hydrogen column density in the MCMC model.
This difference in column densities leads to a systematic difference of the absolute abundances (related to hydrogen) for the metal elements.
This bias is proportional to the total hydrogen column, so it is affected by the \ion{H}{1} and \ion{H}{2} ratio.
Based on our modeling, we found that this ratio is about 5, but could have variation of a factor of 2.
Therefore, the absolute metallicity has an uncertainty of 0.3 dex.
However, this abundance bias will not affect the relative abundance between metal elements, which indicates that the abundance pattern between metals is relatively robust.

The median metallicity of the three volatiles (C, N, and S) is $\rm [X/H] = -0.3$.  Relative to this value, we can see whether the refractory elements have the same metallicity or are lower.  This may occur due to depletion onto grains, where \citet{Sembach:1996aa} show the relative depletions for different types of clouds, including warm diffuse halo clouds.  In such clouds, they show that Fe and Ni are depleted (relative to S) by about $-0.57$ and $-0.77$ dex, with a range of about 0.1 (after their Fig. 6).  We find a similar relative depletion (Table 3) of $-0.5 \pm 0.1$.  The relative depletions given by \citet{Sembach:1996aa} are less for Si ($-0.18$) and Mg ($-0.42$) and are consistent with our relative abundance differences of $-0.4 \pm 0.2$ (Si) and$ -0.6 \pm 0.3$ (Mg).  We conclude that the our absorption line system is similar to warm diffuse Galactic clouds where depletion affects the refractory elements.  The depletion-corrected metallicity appears to be about $\rm [X/H] = -0.3\pm 0.3$.

\begin{table}
\begin{center}
\caption{Photoionization Model and Metallicity}
\label{metallicity}
\begin{tabular}{lccccrr}
\hline\hline
Ion & $\log N$ & $\sigma_N$ & $\log N$ & $\log N$ & $\rm [X/H]^a$& $\rm [X/H]^a$ \\
 & obs. & dex & Ratio & MCMC & Ion & Element\\
\hline\hline
  \ion{H}{1} & 20.06 & 0.10 & 20.06 & 20.06 & $...$ & $...$ \\
  \ion{H}{2} & $...$ & $...$ & 20.68 & 21.01 & $...$ & $...$ \\
   \ion{C}{1} & 13.76  & 0.17 & 14.04 & 14.49 & $-0.28$ & $-0.3 \pm 0.5^b$\\
  \ion{C}{2} & 15.8  & $>$ & 17.15 & 17.43 & $...$ \\
   \ion{N}{1} & 15.71  & 0.10 & 15.91 & 16.02 & $-0.20$ & $-0.2 \pm 0.2$\\
   \ion{N}{5} & 13.97  & 0.10 & 13.07 &  12.86 & $0.90^c$ \\
  \ion{Mg}{1} & 12.76  & 0.24 & 14.03 & 14.32 & $-1.27$ & $-0.9 \pm 0.4$\\
 \ion{Mg}{2} & 15.3  & 0.3 & 15.86 & 16.13 & $-0.6$ \\
 \ion{Si}{2} & 14.8  & $>$ & 16.13 & 16.34 & $...$ & $-0.7 \pm 0.2$\\
\ion{Si}{3} & 14.7  & $>$ & 15.79 & 16.12 & $...$ \\
 \ion{Si}{4} & 14.11  & 0.04 & 14.80 & 14.69 & $-0.69$ \\
  \ion{P}{2} & 13.76  & 0.09 & 14.13 & 14.41 & $-0.37$ & $-0.4 \pm 0.2$\\
  \ion{S}{2} & 15.33  & 0.02 & 15.73 & 16.06 & $-0.40$ & $-0.4 \pm 0.2$\\
 \ion{Fe}{2} & 15.10  & 0.04 & 15.88 & 15.96 & $-0.78$ & $-0.8 \pm 0.2$\\
 \ion{Ni}{2} & 14.19  & 0.06 & 14.97 & 15.24 & $-0.78$ & $-0.8 \pm 0.2$\\
\hline
\end{tabular}
\end{center}
$^a$ The relative abundances of the ions are calculated using the ratio-matched model. The uncertainty in the photoionization model is assumed to be 0.2 dex.\\
$^b$ The abundance of carbon is derived from \ion{C}{1}, but the carbon amount is dominated by \ion{C}{2}. Therefore, we use the uncertainty from the \ion{C}{2} fitting.\\
$^c$ \ion{N}{5} is not from the same phase of the \ion{N}{1}, so we ignore this ion when calculating the nitrogen abundance.
\end{table}

\begin{figure*}
\begin{center}
\includegraphics[width=0.47\textwidth]{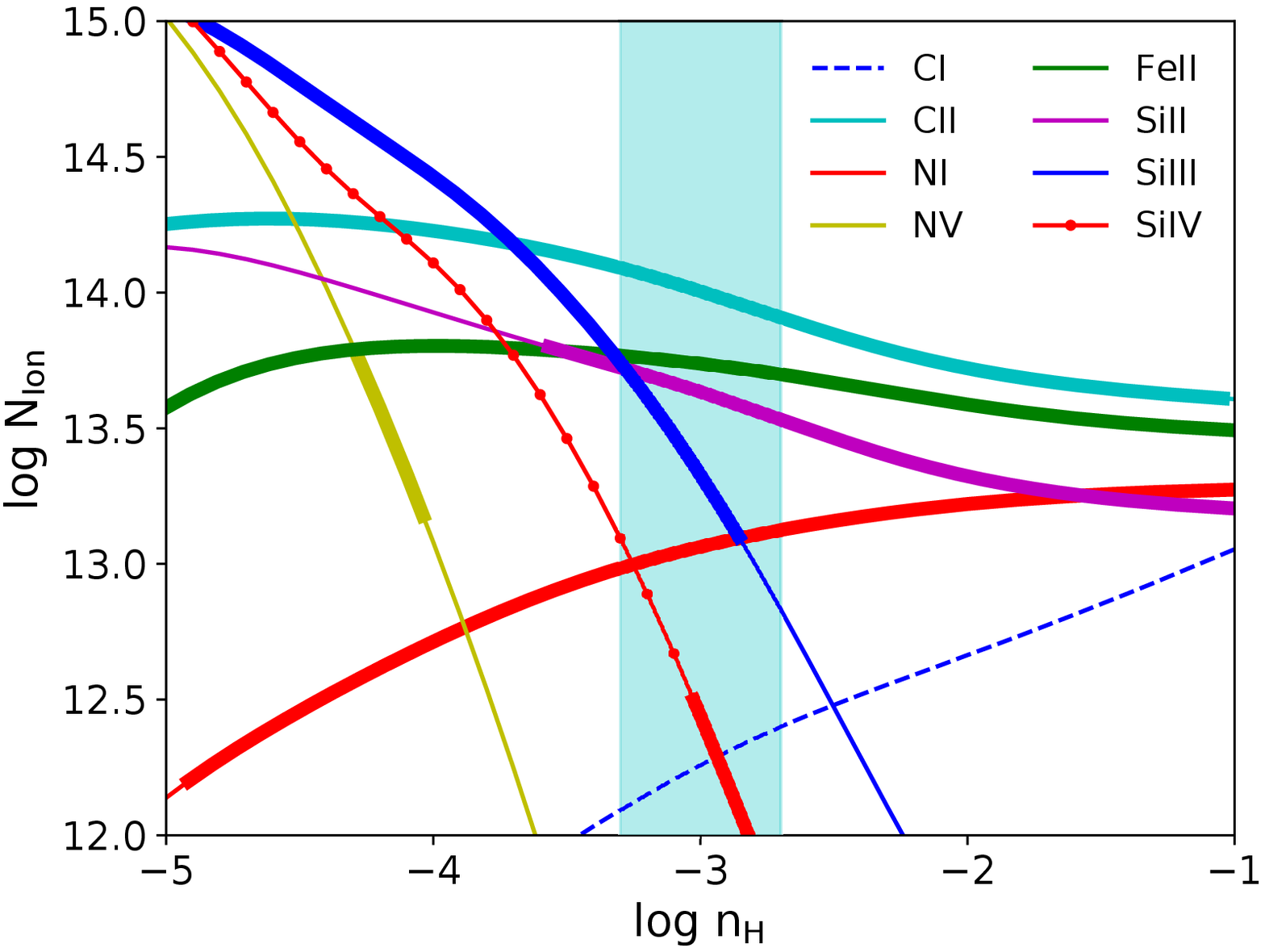}
~
\includegraphics[width=0.47\textwidth]{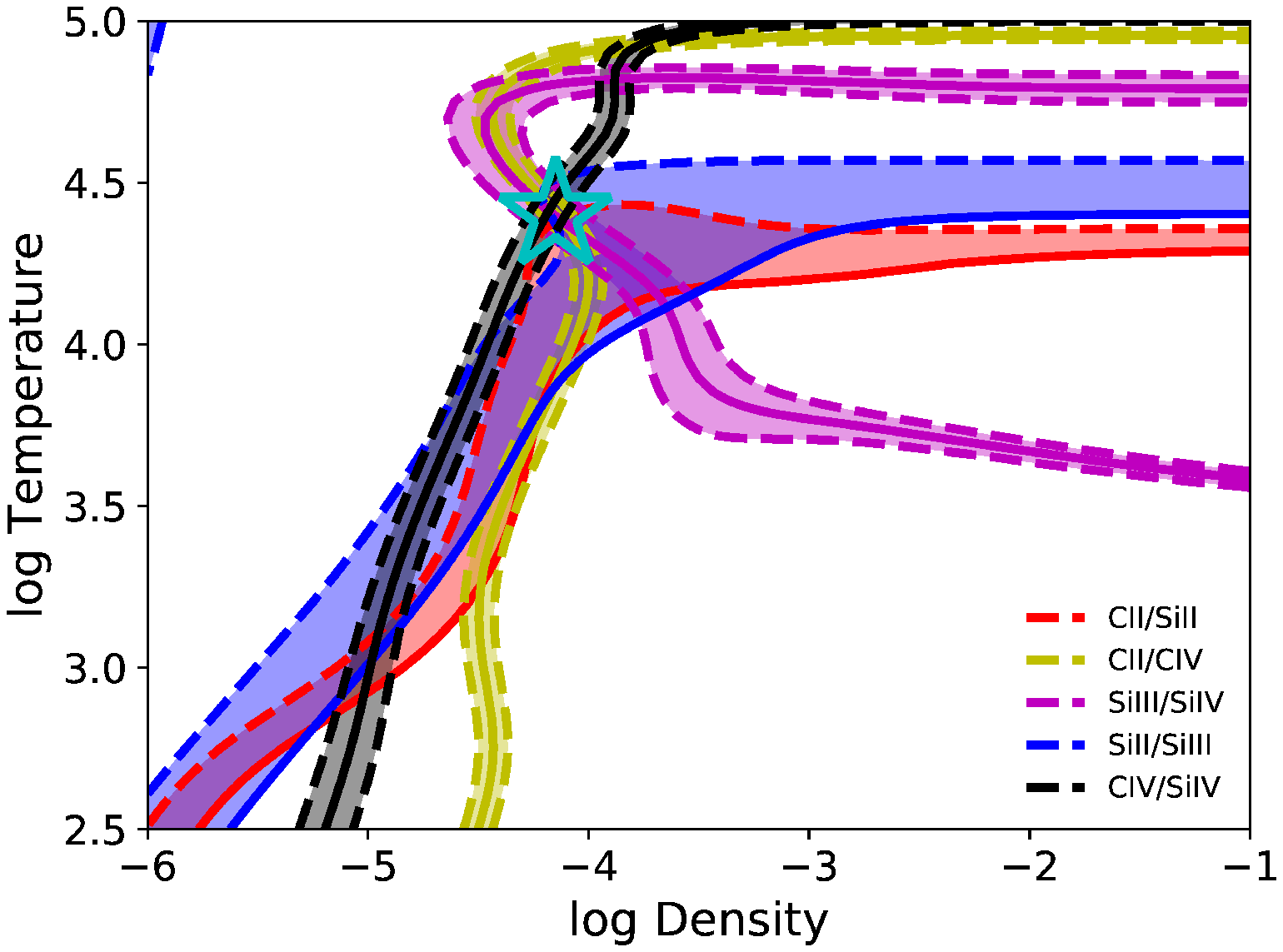}
\end{center}
\caption{{\it Left panel}: the photoionization model of the HVC at $v=+100\kms$ with $Z=Z_\odot$. The \ion{Si}{4} and \ion{N}{1} are upper limit measurements, while the others are detections. The model prediction is in the thin line, while the thick line is the observation constraint. Most ions can be reproduced in the phase of $\log n_{\rm H} = -3.0$. In this model, the relative abundances are $\rm [C/H] = -1.9$, $\rm [Si/H] = -1.3$ and $\rm [Fe/H] = -1.1$. {\it Right panel}: the solution for the system toward 3C 66A without radiation transfer. The solution is $\log T = 4.40\pm0.05$ and $\log n_{\rm H} = -4.15\pm 0.05$. The solution with radiation transfer agrees with this solution, so this is the final solution for this system. The relative abundance between carbon and silicon are solar \citep{Asplund:2009aa}. Since the \ion{Si}{2} column density only has a upper limit, the uncertainty is one-sided. The colored regions are the $1\sigma$ uncertainty regions.}
\label{hvc_model}
\end{figure*}

\subsubsection{\ion{C}{2} Density Sensitive Line}
The \ion{C}{2}* $\lambda 1335.7\rm~\AA$ is absorbed from an excited level, which means the gas density should be high enough to collisionally populate the lower level.
Therefore, the strength of this line can be used to determine the density.
In collisional ionization equilibrium (CIE),
\begin{equation}
\frac{N({\rm C~ II}*)}{N({\rm C~ II})} = \frac{n_{\rm e} k_{01}}{n_{\rm e} k_{10}+A_{10}},
\end{equation}
where $k_{01}$ and $k_{10}$ are the upward and downward collisional rate coefficients, and $A_{10}$ is the spontaneous decay rate.
Then, the \ion{C}{2}*/\ion{C}{2} ratio has a dependence on the density ($n_{\rm e}$) and the temperature (the temperature is in the terms $k_{01}$ and $k_{10}$).

We used the atomic data from CHIANTI to calculate the level populations at different temperatures and densities \citep{Del-Zanna:2015aa}.
Since the \ion{C}{2}* cannot be photoionized, radiation transfer is ignored and the ratio between level populations in CIE is the observed column density ratio (Fig. \ref{model_major}).

Since the \ion{C}{2} column density cannot be well constrained, we adopt the prediction of the photoionization model as $\log N =16.8$, and the uncertainty is adopted from the direct Voigt profile fitting as $0.5 \rm~dex$.
With the \ion{C}{2}* column $\log N = 14.14\pm0.04$, the adopted $\log N({\rm CII})/N({\rm CII})$ ratio is $-2.7\pm 0.5$.
As described in the last section, the preferred temperature is  $\log T = 4.22\pm 0.04$, so the expected density is $\log n_{\rm H} = -1.26 \pm 0.51$.

Combined with the ionization parameter derived from the photoionization model, the strength of the incident field is determined to be 150 times larger than the UVB.
This value is consistent with the escaping light at $5\rm~kpc$ for the MW, which is derived from the OB stars \citep{Fox:2005aa}.

In summary, we find that the absorption system at $v=-30\kms$ has a density of $\log n_{\rm H} = -1.26 \pm 0.51$, and the fixed-temperature photoionization model suggests a temperature of $\log T=4.22\pm 0.04$ and a total hydrogen column density of $\log N_{\rm H} = 20.81 \pm 0.20$. 
Then, the path length of the absorbing gas is $3.8_{-2.6}^{+8.5}\rm~kpc$ combining the measured density and the model hydrogen column density.

\subsubsection{Uncertainties in the Photoionization Model}
We made several assumptions to build the photoionization model, which may lead to uncertainties of the derived physical parameters.
First, we assumed that the \ion{H}{1} column density from the 21 cm line corresponds to the UV absorption system, although the UV line of sight is a pencil beam while the \ion{H}{1} emission is from an approximately $1'$ beam.
If the gas is rotating, the two components around $0 \kms$ seen in the \ion{H}{1} 21 cm line are from different sides of the minor axis, while the UV absorption features will only show one component in the red or blue side.
However, as discussed in the section 4.1, the line width is not dominated by the rotation, which indicates the line profiles should be similar at both sides of the minor axis.
Also, Fig. \ref{Lya} shows the \ion{H}{1} 21 cm line is consistent with the Ly$\alpha$ feature.

Second, we assumed that the temperature is a constant, ignoring the radiation transfer.
In our model, we varied the temperature rather than adopting the photoionization temperature from the Cloudy calculation because of two reasons.
First, The photoionization temperature in the our model is incorrect, where we employ the UVB to obtain the ionization parameter, and increase the density and the incident field correspondingly to account for the escaping flux.
This increase could keep the same ionization parameter, but will break the balance between the cooling and the heating, which have different dependence on the density.
Therefore, the calculated photoionization temperature in our model is not the true temperature when we correct the escaping flux.
Second, as discussed in Section 4.1, the dynamics of this gas is dominated by the turbulence or the outflow, which indicates the gas is not in a quiescent state.
These mechanisms introduce additional heating sources (e.g., the shocks), which raises the photoionization temperature.
Therefore, in our model, the temperature is a variable that need to be inferred.

Third, we assumed that the shape of the incident radiation field could be approximated by the cosmic UVB.
For NGC 891, although the spectrum of the escaping light is not known, studies of the diffuse ionized gas indicate that the expected incident field is more intense and harder than from the MW, shown by the high helium ionization fraction ($\approx 70\%$) and the high [\ion{Ne}{3}]/[\ion{Ne}{2}] ratio \citep{Rand:1997aa, Rand:2008aa}.
Therefore, the escaping flux of NGC 891 may be different from the UVB, since the MW escaping light has a similar shape to the UVB \citep{Fox:2005aa}.

Here, we consider whether the variation of the incident light can affect our modeling of low ionization state ions.
The ionization fraction is dominated by the temperature and the incident field strength near the ionization potential.
For low and intermediate ionization state ions (lower than \ion{Si}{4}), the escaping light should have a similar shape (at low energies) to the UVB, which is dominated by star light.
Modeling of the NGC 891 diffuse ionized halo also suggested that low ionization metal lines (lower than \ion{Ne}{3}; 63 eV) can be modeled by a photoionization model with escaping star light \citep{Rand:2008aa}.
The temperature in our model is not adopted from the photoionization model, thus, it is not affected by the incident field.
Therefore, we suggest that our modeling should not be affected significantly by the variation of the incident field shape with the similar low energy radiation.

The shape of the UVB also leads to variations in the metallicity \citep{Zahedy:2019aa}.
This is mainly affected by the relative radiation strength between HI and metal ions.
Roughly, the UVB shapes are approximated as power laws with a constant slope at low energies (lower than \ion{Si}{4}) in different UVB models \citep{Haardt:2012aa, Khaire:2015ab}.
Therefore, the uncertainty of the slope leads to biases of relative abundance measurements for different elements in one system.
The absolute abundances could be increased or decreased systematically, while the relative abundances between metals should not change significantly.

\subsection{The High Velocity Cloud at $v=100\kms$ toward LQAC 035+042 003}
\label{hvcModel}
We built a pure photoionization model with $Z=Z_\odot$ for this HVC using the UVB-only radiation field.
The \ion{H}{1} column density is adopted for the $v=114\kms$ component, while UV ions are around $v=90-110\kms$.
As shown in Fig. \ref{hvc_model}, we find a solution with $\log n_{\rm H} = -3.0 \pm 0.3$ ($\log U = -3.4 \pm 0.3$), which is determined by the three silicon ions. 
This solution is driven mainly due to the overlap between the lower limit of \ion{Si}{3} and the upper limits of \ion{Si}{2} and \ion{Si}{4}.
Therefore, the \ion{Si}{3} uncertainty caused by the AGN outflow \ion{Ne}{6} will not affect the photoionization model.
In this model, the metal abundances are $\rm [C/H] = -1.9$, $\rm [Si/H] = -1.4$ and $\rm [Fe/H] = -1.1$.
The total hydrogen column density is $\log N_{\rm H} = 19.71 \pm 0.36$.

The upper limit of \ion{C}{2}* is $\log N < 13.2$ for a $2 \sigma$ constraint assuming $b = 20\kms$. 
This implies a density of $\log n_{\rm H} < 1 \cc$, which is much larger than the preferred density in the photoionization model.
Similar to the absorption system at $v=-30\kms$, the incident radiation field is larger than the UVB-only field.
Assuming the incident field is 10 times larger, the density will be 10 times larger correspondingly, hence the path length will also be 10 times smaller ($1.7_{-1.0}^{+2.5}\rm~ kpc$).
From Section \ref{major_absorption}, it is known that the star light is 150 times more intense than the UVB at 5 kpc.
A 10 times larger incident field would imply a distance of $19\rm~kpc$ assuming the radiation is isotropic in all directions.
If the distance is $10~\rm kpc$ (typical distances for the MW's HVCs; \citealt{Wakker:2008aa}), the path length will be $0.45^{+0.67}_{-0.27} ~\rm kpc$.

In the preferred model, most of the detected ions and the upper limit measurements are reproduced in the same phase, except for \ion{C}{1} and \ion{N}{5}.
The very weak \ion{C}{1} ($\approx 2 \sigma$) cannot be in the same phase as the much stronger line of \ion{C}{2}.
Also, the velocity of \ion{C}{1} ($81\kms$) is $\approx 20\kms$ offset from the mean velocity of other ions ($100\kms$), which is a $\approx 2 \sigma$ difference.
Therefore, we suggest that the offset \ion{C}{1} at $v=81\kms$ might be contaminated by other systems (e.g., a weak Ly$\alpha$ system).
The high ionization state ion \ion{N}{5} is not in the same phase of the low ionization state ions (i.e., lower than \ion{Si}{4}) and may be related to cooling or mixing between the hot X-ray gas and the warm component.

\subsection{The Cloud at $v=30\kms$ toward 3C 66A}
For this system, the Cloudy model is not well-constrained, since there are no precise measurements on the \ion{H}{1} column density.
Therefore, we try to find a solution, initially ignoring radiation transfer effects.
The ionization fractions from \citet{Oppenheimer:2013aa} are adopted, which include the UVB-only radiation and assume ionization equilibrium.
As shown in the right panel of Fig. \ref{hvc_model}, the preferred solution is $\log T=4.40\pm 0.05$ and $\log n_{\rm H} = -4.15 \pm 0.05$.
In this model, the required total hydrogen column density is $\log N = 18.06$ assuming solar metallicity.

Using these values, we run a Cloudy model to determine the effect of radiation transfer.
In this model, we fix the metallicity at the solar metallicity, the temperature at $\log T = 4.40$, and the density at $\log n_{\rm H} = -4.15$.
Adding the radiation transfer does not affect the predicted column densities of all ions (Table \ref{3c_model}).
We also consider a model with lower metallicity $Z = 0.01 Z_\odot$ (and $\log N({\rm H}) = 20.06$), which predicts a higher \ion{H}{1} and similar column densities of other ions compared to the solar metallicity model.

The path length of this system varies from $5.26^{+0.64}_{-0.57}\rm~kpc$ to $526^{+64}_{-57}\rm~kpc$ for the metallicity of $Z_\odot$ to $0.01 Z_\odot$.
Since the impact parameter is $108 \rm~kpc$, the strength of the galaxy radiation field is about half of the UVB, adopting 150 times the UVB at $5\rm~ kpc$ (Section 3.1.2) and spherical symmetry.
Then, the path length is reduced by a factor of 1.5 due to the increased incident field.
The metallicity is unlikely to be as low as $0.01Z_\odot$, which leads to a path length larger than the virial radius of NGC 891.
If we adopt the extended hot gas metallicity of $0.14 Z_\odot$ \citep{Hodges-Kluck:2018aa}, the path length is around $20 - 40 \rm~kpc$.

\begin{table}
\tablewidth{1.0\columnwidth}
\begin{center}
\caption{Ionization Models for the 3C 66A System}
\label{3c_model}
\begin{tabular}{cccccc}
\hline
\hline
Ion & $\log N$ & $\sigma_N$ & $\log N_{\rm m}$ & $\log N_{\rm m}$ &  $\log N_{\rm m}$ \\
& obs. & dex & OS13 & $\rm [Z/H] = 0$ & $\rm [Z/H] = -2$ \\
\hline
	\ion{H}{1}  &  19.5  & $<$   &  14.91 &  14.92  &  16.84 \\
   \ion{C}{2}   &  13.03  &  0.11   &  13.03  &  13.02   &  13.00 \\
   \ion{C}{4}   &  13.79  &  0.03   &  13.81  &  13.81  &  13.78 \\
\ion{Si}{2}   &  $11.9$  &  $<$   &  11.50  & 11.42  &  11.39 \\
\ion{Si}{3}   &  12.70  &  0.05   &  12.79  &  12.71  &  12.69 \\
 \ion{Si}{4}   &  12.45  &  0.10   &  12.53  &  12.55  &  12.54 \\
\hline

\end{tabular}
\end{center}
\end{table}

\section{Discussion}

\subsection{The Major System in LQAC 035+042 003}
This work presented UV spectra toward a quasar (LQAC 035+042 003) that is projected behind the edge-on galaxy NGC 891 at a height of 5 kpc from the midplane and near the minor axis. A previous {\it HST}/STIS observation, with the G230L grating, detected absorption lines at the systemic redshift of NGC 891, notably from the ions \ion{Fe}{2} and \ion{Mg}{2}.  The resolution of that observation was $400 \kms$ (FWHM) at $2400\rm~ \AA$.  The COS FUV G130M spectrum presented here improves on the spectral resolution by about a factor of 25 ($15 \kms$ ), which has been crucial in identifying lines and resolving individual velocity components.  Also, the COS spectrum includes lines from elements that are usually not strongly depleted onto grains (C, N, S) as well as elements that, under the right conditions, can be strongly depleted onto grains (Fe, Ni, Si).  By measuring the abundance of this range of elements, we obtain the metallicity of the absorbing gas and the degree of depletion, which appears to be low.

Two absorption systems were identified toward LQAC 035+042 003, a broad set of lines from the warm ionized gaseous halo and a lower column cloud that is separated from the main emission by about $110 \kms$, referred to here as the high velocity cloud (NGC 891 HVC).  Columns for 13 ions are obtained for the primary absorption, and when modeled with CLOUDY, we obtain a consistent fit with log$T = 4.22 \pm 0.04$, log$n_{\rm H} = -1.25 \pm 0.51$, log$N_{\rm H} = 20.81 \pm 0.20$, and a characteristic size for the absorbing region of $\approx 4$ kpc.  The metallicity was obtained for eight elements (Table \ref{metallicity}).

For this warm gas, we obtained a metallicity of $\rm [X/H] = -0.3\pm 0.3$, which is higher than that obtained from the X-ray emitting gas.  The metallicity of the $kT = 0.2\rm~ keV$ hot halo is $Z = 0.14^{+0.09}_{-0.04} Z_\odot$ ($\rm [X/H] = -0.85$ for both O and Fe; \citealt{Hodges-Kluck:2018aa}), about three times lower than the half solar ($\rm [X/H] = -0.3$) that we find in the absorption system. These two metallicities differ at the 2 $\sigma$ level, so the difference is significant. Metallicity values can be a tracer for the origin of the gas, and one might posit that a hot extended halo is lower metallicity than gas rising up from the disk, which should have a near-solar metallicity.  Within this construct, we would identify the X-ray emitting gas as part of the hot extended halo that is accreting onto the galaxy disk. Then, the $10^4\rm~ K$ UV absorbing gas would be understood as a mixture of the disk gas and the hot X-ray emitting medium.  Interaction and mixing between the two components has been proposed \citep{Fraternali:2017aa}, although it is not clear whether this is a viable explanation in detail, as the hot halo mass ($\approx2.4\times 10^8~M_\odot$; \citealt{Hodges-Kluck:2013aa, Hodges-Kluck:2018aa}) is about order of magnitude less than the neutral and warm ionized gas for scale height $< 10$ kpc.

The observed UV absorption system could have three origins: infall from beyond the virial radius; outflow from the disk; and a  rotating disk.
First, infall is not preferred due to the relatively high metallicity of $\approx 0.5 Z_\odot$.
Then, to distinguish between outflow and a rotating disk, we consider the dynamical information from the \ion{H}{1} 21 cm line and UV metal lines, since the dynamics of observed gases are dominated by the non-thermal broadening, as the thermal broadening at $\log T\approx4.2$ is only around $16\kms$.
The 21 cm line and the UV lines have similar velocity ranges, lying between $-100\kms$ and $50\kms$, and with $b$ factors of $\approx 50 \kms$ (the  single component fitting of \ion{H}{1} line in \citealt{Bregman:2013aa} and Table \ref{NGC891}).
This favors the outflow model rather than the rotating disk.
According to \citet{Oosterloo:2007aa}, the rotation curve at $z=4.5\rm~ kpc$ has a linear dependence on the radius until $R\approx8\rm~ kpc$.
With this rotation curve, the rotation velocity dispersion is small at a given projected radius, while the velocity offsets  are different at different projected radii.
If the line widths are dominated by the rotation, the \ion{H}{1} 21 cm line should have a larger $b$ factor than the UV lines because of the lower spatial resolution (the large beam size) of radio observation.
However, the opposite occurs.
Therefore, we conclude that the velocity dispersion of this absorption system is non-thermal and not due to rotation, and an additional broadening mechanism is required to reproduce the observations around the galactic center.
This is consistent with the studies of extended diffuse ionized gas \citep{Boettcher:2016aa}, which showed the H$\alpha$ emission line profile favors a ring model with an inner boundary of $\gtrsim 2\rm~kpc$.
However, current observation cannot uniquely identify the additional mechanism, which might be random turbulence or a biconical outflow.
In the outflow model, the geometry has to be biconical to have the observed radial velocity, since outflows perpendicular to the disk will have a zero radial velocity for edge-on galaxies.
Then, the line widths of different ions are mainly determined by the gas density distribution along the sightline \citep{Fox:2015aa, Savage:2017aa}.

The total hydrogen column is greater than the 21 cm HI column along the same direction, with the ratio being $N({\rm HI + HII})/N({\rm HI}) = 5 - 10$. 
This result is driven by the relatively large columns of singly and doubly ionized elements, mainly C and Si.
If the UV absorption system is from a turbulent gaseous disk, then this hydrogen ionization fraction can apply to the low halo material in general ($z < 10$ kpc), which means most of the gaseous mass in the halo is warm ionized gas.
The \ion{H}{1} halo gas mass is found to be $1.2 \times 10^9~ M_\odot$ \citep{Oosterloo:2007aa}, which is an unusually high value for edge-on galaxies.
The presence of this large amount of gas is already challenging to explain \citep{Hodges-Kluck:2018aa}, which becomes worse if a factor of five increase in the mass applies to the whole halo ($\approx 6\times10^9~M_\odot$).
These warm gases ($\log T \approx 4$) will be accreted onto the disk.
The accretion velocity is assumed to be $v_{\rm ac} = 100 \kms$, and the dynamical timescale will be about $4.9 \times 10^7 \rm~yr$ at $5\rm~ kpc$.
Applying the factor of 5 to the \ion{H}{1} mass, the accretion rate from the warm gases is $1.2 \times 10^2 v^{\rm ac}_{100}~M_\odot~\rm yr^{-1}$, where $v^{\rm ac}_{100}$ is the accretion velocity in the units of $100 \kms$.
This accretion rate is much higher than the current star formation rate (SFR) of NGC 891 ($3.8 ~M_\odot ~ \rm yr^{-1}$; \citealt{Popescu:2004aa}), and will likely elevate the star formation rate in the future, probably causing a disk-wide starburst event.

If this warm gas is a biconical outflow from the galactic center, the hydrogen ionization fraction can only apply to the gas around the galactic center.
Then, the situation is more complicated, so no estimate can be derived for the accretion rate, but these warm gases cannot be buoyant.
Therefore, it is more likely to be a galactic fountain that enhances the gaseous halo.

It is also of interest to consider whether the warm gas is in pressure balance with the hot X-ray emitting gas.
Using the {\it XMM-Newton} and {\it Chandra} spectrum,  the virialized hot halo has an approximate thermal pressure of $n_{\rm H}T = 2.5\times10^3~\rm cm^{-3}~ K$ assuming $n_{\rm H} = n_{\rm e}$ \citep{Hodges-Kluck:2018aa}.
Considering the filling factor, this value is the lower limit of the pressure, since the density is derived from the normalization parameter ($\int n_{\rm e} n_{\rm H} dV$).
The modeling of the UV absorption system leads to a thermal pressure of $n_{\rm H}T = 0.9 \times 10^3 \rm~cm^{-3}~K$.
For the warm gas, the density is obtained from the density sensitive line, which is not affected by the volume filling factor.
Therefore, there is a significant pressure difference between the warm and hot gases.
This difference may be made up by involving other pressure terms, such as the turbulent pressure or the magnetic pressure, since the thermal pressure is found to be non-dominant in the MW interstellar medium \citep{Cox:2005aa}.

\subsection{The HVC in LQAC 035+042 003 -- Cold Mode Accretion?}

We detected an absorption system at $v=640\kms$ ($110 \kms$ relative to NGC 891), which is not necessarily a HVC of NGC 891.
For completeness, it is worth noting that a galaxy group with the central galaxy of NGC 1023 ($v=637 \kms$; \citealt{de-Vaucouleurs:1991aa}) is $0.85~\rm Mpc$ away from LQAC 035+042 003 \citep{Trentham:2009aa}.
Although this galaxy group is closer to the absorption system in velocity, we suggest that it is unlikely to be associated with the galaxy group, because the \ion{H}{1} column is too high ($\log N > 19$) to be typical of the intragroup medium ($\log N \sim 14$; \citealt{Stocke:2014aa}).
Therefore, the absorption system at $640\kms$ is associated with the galaxy NGC 891 rather than large structures beyond the NGC 891 halo.

The modeling of the HVC at $110 \kms$ indicates that it has a metallicity of $\rm [X/H] \approx -1.5$, which is lower than the main body of the absorbing gas ($\rm [X/H] = -0.3$) and the X-ray emitting extended halo ($\rm [X/H] = -0.85$). 
The low metallicity suggests that this gas is accreted from the intergalactic medium (IGM; \citealt{Lehner:2013aa, Wotta:2016aa}), or dwarf galaxies within the NGC 891 halo (e.g., UGC 1807; \citealt{Mapelli:2008aa}).
It may represent a cloud or filament due to the cold mode accretion \citep{Keres:2009aa, Nelson:2013aa}.

Assuming this system is a spherical cloud, the mass of this HVC is $\log M/M_\odot = 5.8$ at a distance of $19 \rm~kpc$ (the radiation field is approximately 10 times the UVB at this radius) or $\log M/M_\odot = 4.6$ at $10\rm ~kpc$.
Then, the accretion rate from this cloud is around $6.8\times 10^{-3} M_\odot \rm ~yr^{-1}$ or $9.2\times 10^{-4} M_\odot~\rm yr^{-1}$ assuming the infall velocity of $200 \kms$. 
The number of such clouds within each radius is estimated assuming the volume filling factor of HVC is $\approx 2\%$ \citep{Richter:2012aa} which leads to $2.0\times10^2$ or $2.8 \times 10^1$ clouds within 10 kpc or 19 kpc. 
Finally, the accretion from the HVC is $0.2M_\odot~\rm~yr^{-1}$ in both cases, which is comparable with the measured accretion rate of the MW HVC ($0.08 M_\odot~\rm yr^{-1}$; \citealt{Putman:2012aa}).
This values set an upper limit of the cold mode accretion rate, since not all of the HVCs have low metallicity.
Another way to estimate the HVC mass of NGC 891 is by assuming the HVC has a covering factor of $50\%$ within a radius of 10 kpc or 19 kpc.
Then, the total mass of the HVC is $\log M = 8.1$ or $\log M= 8.7$ with the assumed radii, which is also comparable with the MW ($\log M = 7.9$; \citealt{Putman:2012aa})

In the MW, studies of sightlines towards the galactic center discovered that HVCs are common \citep{Fox:2015aa, Savage:2017aa}.
The metallicity measurements indicate the disk origin of these gases with $\rm [S/H] = 0.02$ or $\rm [S/H] = 1.38$ \citep{Savage:2017aa}, which are consistent with the outflow from the disk.
In the case of NGC 891, we found the HVC in LQAC 035+042 003 cannot be an outflow due to the low metallicity, but it is not clear whether this system is close to the galactic center.
It is still possible that this system is at the edge of the NGC 891 disk with a small projected distance to the galactic center.

\subsection{The Absorption System in 3C 66A}
The 3C 66A sightline has an impact parameter of 108 kpc, roughly the major axis direction and $21.8^\circ$ above the disk.
This gas is modeled by a photoionization model, with $\log T = 4.40 \pm 0.05$ and $\log n_{\rm H} = -4.15 \pm 0.05$.
Considering the escaping flux from NGC 891, the incident field can be boosted by a factor $\approx 1.5$.
Then the density is increased correspondingly, which leads to a pressure of $3-5\rm~ cm^{-3} ~K$, which is lower than the expected ambient gas pressure ($50-100\rm~ cm^{-3} ~K$) assuming the virial temperature ($\approx 10^6\rm~K$) and the typical density (two hundred times the critical density; $n_{200} \approx 5\times 10^{-5}\cc$).

It is unknown whether this system is assigned to NGC 891, since there are several satellites with smaller projected distances to 3C 66A.
GSC2.3 NCIA030805 is found to have a projected distance of 31.2 kpc from 3C 66A, and 95.5 kpc from NGC 891.
A spectroscopic redshift is not available for this galaxy, but it is assigned to be in the NGC 891 group by its color \citep{Schulz:2014aa}.
Another dwarf galaxy is UGC 1807 (or GSC2.3 NBZ5012371), which is projected 59.8 kpc away from 3C 66A and 75.9 kpc from NGC 891.
This galaxy is believed to be within the virial radius of NGC 891, based on leading interaction features on the \ion{H}{1} disk \citep{Mapelli:2008aa}.

The radial velocity ($583 \kms$) of the UV absorption system is opposite to the rotation of NGC 891.
The gaseous halo is more likely to be co-rotating with the disk out to the half of the virial radius \citep{Ho:2017aa}.
Therefore, this absorption system towards 3C 66 is unlikely to be a part of the extended co-rotating gaseous halo of NGC 891.
Perhaps this gas was stripped from the nearby dwarf or it is accreting onto one of these galaxies from the near side.

\section{Summary}
We analyzed the {\it HST}/COS spectra to detect the gaseous components in the NGC 891 halo either close to the minor axis (LQAC 035+042 003) or further away along the major axis (3C 66A).
Three absorption systems are detected in these two sightlines. 
Our results are summarized as:
\begin{itemize}
\item In the spectrum of LQAC 035+042 003, the primary absorption system near $v=-30\kms$ at $z=0.001761$  is modeled as a gas with $\log T = 4.22\pm 0.04$, $\log n_{\rm H} = -1.26\pm 0.51$, and $\log N_{\rm H} = 20.81\pm 0.20$.
Abundances are measured for eight elements, showing a typical depletion pattern seen in the MW warm diffuse gas.
The volatile elements (C, N, S) show a metallicity of $\rm [X/H] = -0.3\pm0.3$, which suggests that the high-metallicity disk-originating gas is mixed with the virialized hot halo ($Z\approx 0.14 Z_\odot$) seen in the X-rays \citep{Hodges-Kluck:2018aa}.
This leads to a more massive warm gas disk than the \ion{H}{1} disk from 21 cm emission, and along this sightline; the H/\ion{H}{1} ratio is about $5\pm1$.

\item The HVC in LQAC 035+042 003 is separated from the primary system by $\approx + 110 \kms$, with $\log U = -3.4 \pm 0.3$ and the metallicity is lower than $0.1Z_\odot$ with an average value of $\rm [X/H]=-1.5$ for C, Si, and Fe.
This is different from the HVC seen in the MW (about solar metallicity) around the galactic center \citep{Savage:2017aa}, which are about the solar metallicity. 
Therefore, we suggest that this low metallicity HVC in NGC 891 is likely to be an accreted cloud from the more distant halo and has not mixed effectively with the hot halo or the warm gas closer to the disk.

\item The absorption system in the 3C 66A spectrum is suggested to be associated with a satellite around NGC 891 rather with NGC 891 itself, which is mainly because the velocity of this system is opposite to the rotational velocity of the \ion{H}{1} disk.
\end{itemize}

\acknowledgments
We thank Jiangtao Li and the anonymous referee for helpful comments. We gratefully acknowledge support from grants HST-GO-12904 and NNX16AF23G. 

\bibliographystyle{apj}
\bibliography{MissingBaryon}

\begin{figure*}
\begin{center}
\includegraphics[width=0.83\textwidth]{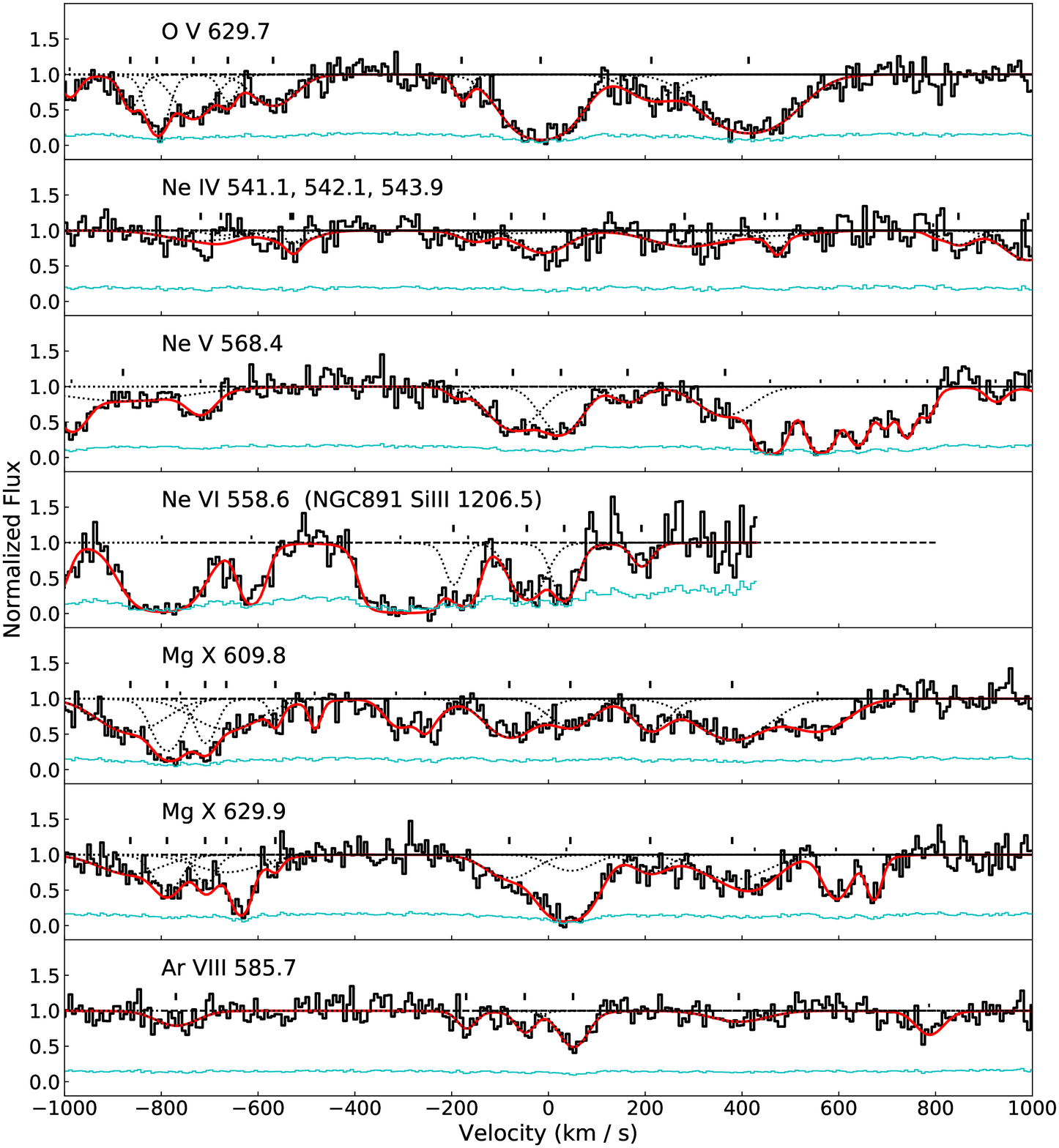}
\end{center}
\caption{The AGN outflow at $z=1.1655$. The red solid lines are the total model, while the black dotted lines are individual components associated with the outflow. The thick long bars indicates the position of the AGN outflow components, while the short bars are other components.}
\label{AGNoutflow}
\end{figure*}

\section*{Appendix}
We match the detected absorption features using the system detection procedure in \citet{Qu:2016aa} to find absorption systems.
Here, we only give a brief summary on this procedure.
First, we apply the Bayesian Block algorithm \citep{Scargle:2013aa} to the total count array to resample the coadded spectrum. 
The absorption features are defined as the low-flux and short-width blocks. 
Then, the detected absorption features are matched using an input line template composed of strong intergalactic medium lines (e.g., \ion{C}{3} $\lambda 977.0\rm~\AA$ and \ion{O}{5} $\lambda 629.7\rm~\AA$). 
After the systems are detected automatically, we double check the systems by eye to confirm the existence.

We detect 220 absorption features (some have blended lines in one feature) in the entire FUV G130M spectrum.
Including the MW ISM and the NGC 891 absorption features, we find eight systems with metal lines, and another four systems with matched Ly series lines.
There are also three Ly$\alpha$-only systems with matched galaxies or galaxy groups (clusters). 
The full list of identified ions are listed in Table \ref{ions}, where all of measurements are from the Voigt profile fitting.
This may lead to large uncertainties for saturated lines, which has been marked in this table.

After identifying absorbing systems, the contamination of the NGC 891 absorption lines can be determined, showing that most NGC 891 lines are not blended with other lines.
However, the \ion{Si}{3} $\lambda 1206.5\rm~\AA$ is contaminated by the \ion{Ne}{6} lines of the AGN outflow at $z=1.1655$.
As shown in Fig. \ref{AGNoutflow}, the AGN outflow occurs in a wide velocity range with multiple components, and three major components are around $-800\kms $, $0 \kms$, and $400 \kms$ at $z=1.1655$.
The \ion{O}{5}, \ion{Ne}{5}, and \ion{Ar}{7} lines indicate a component with $v \approx - 170  \kms$, which is $v=86 \kms$ of the \ion{Si}{3} at $z=0.001761$.
For this component, there is no detectable \ion{Mg}{10}, so it is an intermediate ionization state system.
Therefore, it is possible that this component contains a considerable contamination to the NGC 891 \ion{Si}{3} absorption features.

However, both \ion{Ne}{6} and \ion{Si}{3} have single lines in the current spectral coverage, which means that these two lines cannot be decomposed using other lines from the same ions.
If we ignore the contribution from the \ion{Ne}{6}, the HVC \ion{Si}{3} column density is $\log N = 13.78\pm 0.16$ and the velocity is $v = 80.7\pm 2.9 \kms$, which differs by $2\sigma$ from the other ions.
Therefore, we introduce the \ion{Ne}{6} component around $-170\kms$ into the fitting and fix the position of the HVC \ion{Si}{3} (as \ion{Si}{2} at $v=89.4\kms$) to determine whether is it possible to obtain an improved fit.
We obtain $\log N = 13.37\pm 0.14$ for the HVC and the NGC 891 major absorption is weakly affected, which is adopted in the main text.
The difference in $\chi^2$ between these two fittings is less than 1, which means that both fittings are acceptable.
Although we prefer the fitting with \ion{Ne}{6}, the HVC \ion{Si}{3} column density remains uncertain.

\begin{longtable}{l r r r}
\tablewidth{0.45\textwidth}
\tablecolumns{2}
\tablecaption{The Identified Ion List in LQAC 035+042 003}
\vspace{-0.3cm}
\tablehead{\colhead{Ion} & \colhead{$\log N~(\rm dex)$} & \colhead{$b~ (\kms)$ } & \colhead{$v~ (\kms)$}}
\multicolumn{4}{l}{$z=-0.0001$; The MW ISM}\\
\hline
   \ion{C}{1}  &  $14.19 \pm   0.06$  &  $24.0  \pm  5.5$  & $ 14.8  \pm  3.5$ \\
  \ion{C}{2}$^b$  &  $15.84 \pm   0.58$  &  $32.0  \pm  5.9$  & $-10.6  \pm  1.4$ \\
 \ion{C}{2}*  &  $14.48 \pm   0.05$  &  $28.3  \pm  3.1$  & $ 11.7  \pm  2.0$ \\
   \ion{N}{1}  &  $15.76 \pm   0.16$  &  $10.0^a$  & $-18.7  \pm  1.1$ \\
   \ion{N}{1}  &  $14.42 \pm   0.10$  &  $10.0^a$  & $ 30.0  \pm  1.5$ \\
 \ion{Si}{2}  &  $16.57 \pm   0.08$  &  $17.7  \pm  0.9$  & $ -8.4  \pm  1.4$ \\
\ion{Si}{3}  &  $14.04 \pm   0.05$  &  $67.5  \pm  4.2$  & $ 14.6  \pm  2.4$ \\
 \ion{Si}{4}  &  $13.47 \pm   0.05$  &  $36.9  \pm  5.3$  & $ -6.0  \pm  3.6$ \\
  \ion{P}{2}  &  $14.15 \pm   0.05$  &  $39.5  \pm  6.2$  & $ 27.5  \pm  4.4$ \\
  \ion{S}{2}  &  $15.68 \pm   0.05$  &  $22.0  \pm  1.3$  & $  6.7  \pm  0.9$ \\
 \ion{Fe}{2}  &  $15.00 \pm   0.05$  &  $40.9  \pm  4.2$  & $ -5.0  \pm  3.2$ \\
 \ion{Ni}{2}  &  $14.17 \pm   0.06$  &  $34.3  \pm  7.0$  & $ 17.5  \pm  4.7$ \\

\hline
\multicolumn{4}{l}{$z=0.001761$; NGC 891}\\
\hline
      \ion{C}{1}  &  $13.76  \pm  0.17$  &  $23.5  \pm 15.7$  &  $-34.6  \pm  10.0$ \\
     \ion{C}{2}$^b$  &  $16.34  \pm  0.54$  &  $53.2  \pm  8.5$  &  $-25.0  \pm  2.6$ \\
    \ion{C}{2}*  &  $14.14  \pm  0.05$  &  $33.3  \pm  5.5$  &  $-34.2  \pm  3.6$ \\
      \ion{N}{1}  &  $15.71  \pm  0.10$  &  $34.1  \pm  1.7$  &  $-24.0  \pm  1.3$ \\
      \ion{N}{5}  &  $13.97  \pm  0.10$  &  $42.8  \pm 14.7$  &  $-27.3  \pm  9.4$ \\
    \ion{Si}{2}  &  $15.60  \pm  0.26$  &  $39.2  \pm  2.9$  &  $-39.1  \pm  1.1$ \\
   \ion{Si}{3}$^b$  &  $15.11  \pm  0.24$  &  $40.0^a$  &  $-50.2  \pm  5.3$ \\
    \ion{Si}{4}  &  $13.89  \pm  0.08$  &  $48.1  \pm  8.6$  &  $-69.2  \pm  8.3$ \\
    \ion{Si}{4}  &  $13.62  \pm  0.16$  &  $22.4  \pm  7.7$  &  $ -5.4  \pm  4.3$ \\
    \ion{Si}{4}$^b$  &  $12.93  \pm  0.31$  &  $32.5  \pm 29.4$  &  $ 57.9  \pm 21.8$ \\
     \ion{P}{2}  &  $13.76  \pm  0.09$  &  $25.1  \pm  9.6$  &  $ -4.1  \pm  6.0$ \\
     \ion{S}{2}  &  $15.33  \pm  0.02$  &  $50.9  \pm  3.2$  &  $-25.5  \pm  2.2$ \\
    \ion{Fe}{2}  &  $15.10  \pm  0.04$  &  $55.1  \pm  4.6$  &  $-34.1  \pm  3.2$ \\
    \ion{Ni}{2}  &  $14.19  \pm  0.06$  &  $53.0  \pm 10.4$  &  $-36.1  \pm  6.7$ \\
\hline
\multicolumn{4}{l}{$z=0.001761$; The NGC 891 HVC}\\
\hline
      \ion{C}{1}  &  $13.66   \pm 0.23$  &  $ 7.3  \pm 14.0$  & $ 81.4  \pm  5.7$ \\
     \ion{C}{2}  &  $13.97   \pm 0.18$  &  $10.0^a$  & $107.1  \pm  3.1$ \\
      \ion{N}{5}  &  $13.47   \pm 0.15$  &  $19.7  \pm 12.8$  & $ 99.6  \pm  7.1$ \\
    \ion{Si}{2}  &  $13.50   \pm 0.08$  &  $14.8  \pm  2.1$  & $ 89.4  \pm  1.5$ \\
   \ion{Si}{3}  &  $13.37   \pm 0.14$  &  $20.0^a$  & $ 89.4^a$ \\
    \ion{Fe}{2}  &  $13.77   \pm 0.18$  &  $20.0^a$  & $107.3  \pm 11.0$ \\

\hline
\multicolumn{4}{l}{$z=0.0149$; A Galaxy Group with NGC 912}\\
\hline
     \ion{H}{1}  &  $13.40  \pm  0.10 $ &  $15.0^a$  &  $-1.8  \pm  5.3$ \\
     \ion{H}{1}  &  $13.68  \pm  0.11 $ &  $14.0  \pm  6.0$  &  $43.3  \pm  2.7$ \\
     \ion{H}{1}  &  $13.23  \pm  0.13 $ &  $15.0^a$  &  $86.3  \pm  5.8$ \\

\hline
\multicolumn{4}{l}{$z=0.0166$; A Galaxy Group with NGC 909}\\
\hline
     \ion{H}{1}  &  $13.00  \pm  0.15$  &  $22.9 \pm  14.6$  & $-274.6  \pm  9.0$  \\  
     \ion{H}{1}  &  $13.62  \pm  0.10$  &  $52.0 \pm  13.0$  & $-114.7  \pm  9.6$  \\  
     \ion{H}{1}  &  $13.50  \pm  0.19$  &  $30.0^a$  & $  -3.2  \pm  6.2$  \\
     \ion{H}{1}$^b$  &  $13.64  \pm  0.21$  &  $89.3 \pm  39.3$  & $  64.9  \pm 38.0$  \\ 

\hline
\multicolumn{4}{l}{$z=0.0200$; Galaxy Cluster Abell 347}\\
\hline
     \ion{H}{1}  &  $13.62  \pm  0.06$  & $ 20.7 \pm   4.1$  & $-458.1 \pm   2.5$ \\
     \ion{H}{1}  &  $13.89  \pm  0.07$  & $115.6 \pm  22.8$  & $-190.3 \pm  14.8$ \\
     \ion{H}{1}  &  $14.39  \pm  0.18$  & $ 26.6 \pm   4.5$  & $ -45.2 \pm   1.7$ \\
     \ion{H}{1}  &  $13.07  \pm  0.16$  & $ 10.0^a$  & $  34.9 \pm   5.3$ \\
     \ion{H}{1}  &  $14.24  \pm  0.11$  & $ 25.9 \pm   5.1$  & $  98.9 \pm   2.5$ \\
     \ion{H}{1}  &  $14.10  \pm  0.05$  & $ 41.5 \pm   7.9$  & $ 189.0 \pm   5.0$ \\
     \ion{H}{1}  &  $13.25  \pm  0.16$  & $ 20.0^a$  & $ 288.6 \pm   6.6$ \\
    \ion{C}{2}  &  $13.63  \pm  0.10$  & $ 20.0^a$  & $-446.9 \pm   5.5$ \\   

\hline
\multicolumn{4}{l}{$z=0.0667$; A Galaxy Group}\\
\hline
     \ion{H}{1}  &  $13.22  \pm  0.12$  &  $20.7 \pm  10.4$ &  $-229.6  \pm  6.2$ \\
     \ion{H}{1}  &  $13.62  \pm  0.06$  &  $26.3 \pm   5.8$ &  $-157.6  \pm  3.6$ \\
   
\hline
\multicolumn{4}{l}{$z=0.1080$}\\
\hline
     \ion{H}{1}  &  $13.93  \pm  0.03$  &  $50.9 \pm   4.8$  &  $21.1  \pm  3.4$ \\

\hline
\multicolumn{4}{l}{$z=0.1134$}\\
\hline
     \ion{H}{1}  &  $14.37 \pm   0.03$  &  $55.3  \pm  3.3$ &   $-10.9 \pm   2.4$ \\

\hline
\multicolumn{4}{l}{$z=0.1532$}\\
\hline
     \ion{H}{1}  &  $14.43 \pm   0.07$  &  $24.4  \pm  1.6$  & $ -44.7 \pm   1.4$ \\
     \ion{H}{1}  &  $13.76 \pm   0.05$  &  $24.5  \pm  3.5$  & $  67.9 \pm   2.2$ \\
\hline
\multicolumn{4}{l}{$z=0.1561$}\\
\hline
     \ion{H}{1}  &  $13.46 \pm   0.09$  &  $34.5  \pm  9.5$  & $-165.4 \pm   6.4$ \\
     \ion{H}{1}$^b$  &  $17.10 \pm   0.48$  &  $ 9.0  \pm  4.8$  & $  10.3 \pm  12.5$ \\
     \ion{H}{1}$^b$  &  $14.90 \pm   0.15$  &  $20.2  \pm  3.4$  & $  62.0 \pm   4.9$ \\
     \ion{H}{1}  &  $13.43 \pm   0.08$  &  $35.4  \pm  9.9$  & $ 159.7 \pm   6.1$ \\
\hline
\multicolumn{4}{l}{$z=0.1579$}\\
\hline
     \ion{H}{1}  &  $13.44 \pm   0.08$  &  $56.0  \pm 14.1$  & $  38.9 \pm   9.5$ \\
\hline
\multicolumn{4}{l}{$z=0.1623$}\\
\hline
     \ion{H}{1}  &  $14.72  \pm  0.15$  &  $40.6  \pm  4.0$  &  $43.3  \pm  1.6$ \\
   \ion{C}{3}  &  $13.65  \pm  0.08$  &  $42.4  \pm 13.6$  &  $39.0  \pm  8.5$ \\
   
\hline
\multicolumn{4}{l}{$z=0.2627$}\\
\hline
     \ion{H}{1}  &  $14.62 \pm   0.04$  &  $33.7  \pm  4.2$  &  $-11.2 \pm   2.7$ \\

\hline

\multicolumn{4}{l}{$z=0.4173$}\\
\hline
     \ion{H}{1}  &  $15.40 \pm   0.03$  & $ 31.9 \pm   3.1$  & $ -6.3  \pm  2.3$ \\
     \ion{H}{1}  &  $15.03 \pm   0.06$  & $ 15.0 \pm   2.8$  & $ 53.7  \pm  2.2$ \\
   \ion{C}{3}  &  $13.61 \pm   0.52$  & $  8.9 \pm   4.9$  & $-12.1  \pm  1.6$ \\
   \ion{C}{3}  &  $13.73 \pm   0.05$  & $ 25.2 \pm   3.1$  & $ 61.4  \pm  2.0$ \\
   
\hline
\multicolumn{4}{l}{$z=0.7553$}\\
\hline
   \ion{O}{3}  &  $15.10  \pm  0.15$  & $ 28.6  \pm  5.4$  & $ -2.1 \pm   2.6$ \\
   \ion{O}{3}  &  $15.01  \pm  0.15$  & $ 25.0  \pm  5.2$  & $102.0 \pm   2.1$ \\
   \ion{O}{3}  &  $14.62  \pm  0.06$  & $ 21.4  \pm  5.1$  & $178.7 \pm   2.6$ \\
    \ion{O}{4}  &  $16.64  \pm  0.15$  & $ 20.0^a$  & $  4.8 \pm   1.5$ \\
    \ion{O}{4}  &  $15.08  \pm  0.20$  & $ 23.5  \pm  7.3$  & $100.8 \pm   3.7$ \\
    \ion{O}{4}  &  $15.23  \pm  0.31$  & $ 23.6  \pm  6.5$  & $170.3 \pm   4.0$ \\
    \ion{N}{2}  &  $14.32  \pm  0.15$  & $ 15.0^a$  & $103.3 \pm   6.8$ \\
   \ion{N}{3}  &  $14.19  \pm  0.06$  & $ 17.9  \pm  3.3$  & $106.4 \pm   2.1$ \\
   \ion{N}{3}  &  $13.65  \pm  0.11$  & $ 10.0^a$  & $172.2 \pm   3.6$ \\
    \ion{N}{4}  &  $13.88  \pm  0.04$  & $ 28.7  \pm  3.4$  & $  6.8 \pm   2.3$ \\
    \ion{N}{4}  &  $13.85  \pm  0.25$  & $ 11.6  \pm  4.5$  & $ 94.4 \pm   2.2$ \\
    \ion{N}{4}  &  $13.43  \pm  0.09$  & $ 20.2  \pm  7.8$  & $174.6 \pm   4.5$ \\
    \ion{S}{2}  &  $12.72  \pm  0.12$  & $ 18.8  \pm  9.7$  & $103.4 \pm   5.8$ \\
   \ion{S}{3}  &  $13.62  \pm  0.04$  & $ 33.3  \pm  4.3$  & $100.1 \pm   3.0$ \\
    \ion{S}{4}  &  $13.56  \pm  0.07$  & $ 46.8  \pm 11.3$  & $  7.7 \pm   6.6$ \\
    \ion{S}{4}  &  $13.07  \pm  0.16$  & $ 19.9  \pm 15.6$  & $107.7 \pm   8.0$ \\
    \ion{S}{4}  &  $13.32  \pm  0.10$  & $ 25.8  \pm 11.1$  & $190.6 \pm   6.5$ \\
     \ion{S}{5}  &  $12.92  \pm  0.10$  & $ 31.5  \pm 10.6$  & $  1.7 \pm   6.9$ \\
     \ion{S}{5}  &  $12.60  \pm  0.17$  & $ 20.0^a$  & $101.6 \pm   9.8$ \\
     \ion{S}{5}  &  $12.86  \pm  0.14$  & $ 42.4  \pm 19.5$  & $188.1 \pm  12.2$ \\

\hline
\multicolumn{4}{l}{$z=1.1655$; The AGN Outflow}\\
\hline
     \ion{O}{5}  &  $13.58 \pm   0.09$  & $ 20.0^a$ & $-862.7 \pm   4.8$ \\
     \ion{O}{5}  &  $14.13 \pm   0.08$  & $ 20.6 \pm   5.1$ & $-808.2 \pm   3.1$ \\
     \ion{O}{5}  &  $14.00 \pm   0.13$  & $ 44.1 \pm  18.6$ & $-732.8 \pm   6.5$ \\
     \ion{O}{5}  &  $13.43 \pm   0.27$  & $ 18.0 \pm  12.3$ & $-661.5 \pm   7.4$ \\
     \ion{O}{5}  &  $13.85 \pm   0.06$  & $ 53.8 \pm  10.9$ & $-568.4 \pm   7.0$ \\
     \ion{O}{5}  &  $13.33 \pm   0.12$  & $ 15.5 \pm   8.1$ & $-179.4 \pm   4.7$ \\
     \ion{O}{5}  &  $14.65 \pm   0.02$  & $ 75.5 \pm   3.7$ & $ -16.1 \pm   2.4$ \\
     \ion{O}{5}  &  $13.77 \pm   0.10$  & $ 62.4 \pm  16.4$ & $ 212.7 \pm  11.3$ \\
     \ion{O}{5}  &  $14.59 \pm   0.02$  & $101.5 \pm   6.2$ & $ 413.9 \pm   4.3$ \\
   \ion{Ne}{4}  &  $14.25 \pm   0.10$  & $ 94.9 \pm  26.0$ & $-717.6 \pm  17.7$ \\
   \ion{Ne}{4}  &  $13.80 \pm   0.15$  & $ 13.4 \pm  11.7$ & $-527.3 \pm   6.0$ \\
   \ion{Ne}{4}  &  $13.86 \pm   0.17$  & $ 36.5 \pm  21.4$ & $-152.7 \pm  13.5$ \\
   \ion{Ne}{4}  &  $14.41 \pm   0.06$  & $ 58.6 \pm  11.4$ & $  -8.9 \pm   7.2$ \\
   \ion{Ne}{4}$^b$  &  $14.01 \pm   0.19$  & $106.4 \pm  58.4$ & $ 447.6 \pm  39.4$ \\
    \ion{Ne}{5}$^b$  &  $14.63 \pm   0.09$  & $ 172.0\pm  42.2$ & $-877.8 \pm  30.7$ \\
    \ion{Ne}{5}  &  $13.66 \pm   0.30$  & $20.0^a$ & $-189.8 \pm  16.2$ \\
    \ion{Ne}{5}  &  $14.82 \pm   0.16$  & $60.1 \pm   19.7$ & $ -73.4 \pm   17.7$ \\
    \ion{Ne}{5}  &  $14.80 \pm   0.16$  & $47.6 \pm   12.8$ & $   25.8 \pm   12.5$ \\
    \ion{Ne}{5}  &  $14.06 \pm   0.16$  & $ 36.47 \pm  19.0$ & $ 163.5 \pm  11.7$ \\
    \ion{Ne}{5}  &  $14.61 \pm   0.08$  & $ 68.4 \pm  16.9$ & $ 365.2 \pm  10.9$ \\
   \ion{Ne}{6}  &  $14.57 \pm   0.25$  & $ 16.7 \pm   18.1$ & $ -196.2 \pm   10.5$ \\
   \ion{Ne}{6}  &  $14.99 \pm   0.07$  & $ 37.2 \pm   9.7$ & $ -44.6 \pm   6.6$ \\
   \ion{Ne}{6}  &  $14.92 \pm   0.11$  & $ 23.9 \pm   7.3$ & $  32.7 \pm   5.5$ \\
   \ion{Ne}{6}  &  $14.23 \pm   0.18$  & $ 22.3 \pm  17.8$ & $ 192.3 \pm  10.2$ \\
    \ion{Mg}{10}$^b$  &  $14.86 \pm   0.21$  & $ 82.6 \pm  29.5$ & $-862.6 \pm  33.9$ \\
    \ion{Mg}{10}  &  $14.88 \pm   0.25$  & $ 32.7 \pm  10.7$ & $-787.2 \pm   6.1$ \\
    \ion{Mg}{10}  &  $14.63 \pm   0.36$  & $ 22.1 \pm  11.4$ & $-708.4 \pm   4.1$ \\
    \ion{Mg}{10}$^b$  &  $14.83 \pm   0.44$  & $ 86.1 \pm  58.8$ & $-664.9 \pm  70.2$ \\
    \ion{Mg}{10}  &  $14.03 \pm   0.23$  & $ 11.0 \pm  11.5$ & $-563.7 \pm   4.8$ \\
    \ion{Mg}{10}  &  $14.84 \pm   0.07$  & $ 61.4 \pm  11.4$ & $ -80.8 \pm   9.1$ \\
    \ion{Mg}{10}$^b$  &  $14.61 \pm   0.12$  & $ 55.2 \pm  17.3$ & $  45.2 \pm  12.8$ \\
    \ion{Mg}{10}  &  $14.58 \pm   0.08$  & $ 43.7 \pm   9.4$ & $ 210.3 \pm   6.3$ \\
    \ion{Mg}{10}  &  $15.03 \pm   0.07$  & $ 89.2 \pm  17.5$ & $ 379.8 \pm   9.3$ \\
  \ion{Ar}{7}  &  $13.12 \pm   0.11$  & $ 53.3 \pm  18.3$ & $-768.8 \pm  12.0$ \\
  \ion{Ar}{7}  &  $12.90 \pm   0.12$  & $ 20.0^a$ & $-169.8 \pm   6.9$ \\
  \ion{Ar}{7}  &  $13.01 \pm   0.11$  & $ 21.8 \pm   9.7$ & $ -49.0 \pm   5.9$ \\
  \ion{Ar}{7}  &  $13.44 \pm   0.05$  & $ 32.6 \pm   5.6$ & $  50.9 \pm   3.7$ \\
  \ion{Ar}{7}$^b$  &  $13.10 \pm   0.13$  & $ 74.2 \pm  27.4$ & $ 393.5 \pm  18.9$ \\

\hline
\multicolumn{4}{l}{$z=1.1788$; The AGN Narrow Line Absorption}\\
\hline
    \ion{O}{4}  &  $14.72  \pm  0.04$  &  $24.8  \pm  2.0$  & $ -9.2  \pm  1.5$ \\
     \ion{O}{5}  &  $14.45  \pm  0.16$  &  $22.5  \pm  3.3$  & $ -8.2  \pm  1.6$ \\
   \ion{Ne}{4}  &  $14.26  \pm  0.06$  &  $22.0  \pm  5.4$  & $-14.1  \pm  3.5$ \\
    \ion{Ne}{5}  &  $14.63  \pm  0.27$  &  $10.8  \pm  4.2$  & $-22.6  \pm  2.0$ \\
\hline
\hline

\label{ions}
\end{longtable}
$^a$ These values are fixed to reduce the uncertainty of the fittings.\\
$^b$ All of the measurements in this table are from the Voigt profile fitting. The marked ions suffer from the large measurement uncertainty due to the saturation of isolated lines or uncertain line centroids in blended situations.\\

\end{document}